\documentclass[useAMS,usenatbib,a4paper]{mn2e}
\include{graphicx}
\title[NIR interline background with FBG OH suppression]{The nature of the near-infrared interline sky background using fibre Bragg grating OH suppression}
\author[Trinh et al.]{Christopher Q. Trinh$^{1}$\thanks{E-mail:
c.trinh@physics.usyd.edu.au}, 
Simon C. Ellis$^{2,1}$, Joss Bland-Hawthorn$^{1,3}$, Anthony J. Horton$^{2}$, 
\newauthor Jon S. Lawrence$^{2,4}$ and Sergio G. Leon-Saval$^{3}$ \\
$^{1}$Sydney Institute for Astronomy, School of Physics, The University of Sydney, NSW 2006, Australia\\
$^{2}$Australian Astronomical Observatory, 105 Delhi Road, North Ryde, P.O. Box 915, NSW 1670, Australia\\
$^{3}$Institute of Photonics and Optical Science, School of Physics, The University of Sydney, NSW 2006, Australia\\
$^{4}$Department of Physics and Astronomy, Macquarie University, NSW 2109, Australia}
\begin{document}

\date{Accepted Received}

\pagerange{\pageref{firstpage}--\pageref{lastpage}} \pubyear{2012}

\maketitle

\label{firstpage}

\begin{abstract}
We analyse the near-infrared interline sky background, OH and O$_{2}$ emission in 19 hours of $H$ band observations with the GNOSIS OH suppression unit and the IRIS2 spectrograph at the 3.9-metre Anglo-Australian Telescope. We find that the temporal behaviour of OH emission is best described by a gradual decrease during the first half of the night followed by a gradual increase during the second half of the night following the behaviour of the solar elevation angle. We measure the interline background at 1.520\,$\micron$ where the instrumental thermal background is very low and study its variation with zenith distance, time after sunset, ecliptic latitude, lunar zenith angle and lunar distance to determine the presence of non-thermal atmospheric emission, zodiacal scattered light and scattered moonlight. Zodiacal scattered light is too faint to be detected in the summed observations. Scattered moonlight due to Mie scattering by atmospheric aerosols is seen at small lunar distances ($\rho\la$\,11\,deg), but is otherwise too faint to detect. Except at very small lunar distances the interline background at a resolving power of $R\approx2400$ when using OH suppression fibres is dominated by a non-thermal atmospheric source with a temporal behaviour that resembles atmospheric OH emission suggesting that the interline background contains instrumentally-scattered OH. However, the interline background dims more rapidly than OH early in the night suggesting contributions from rapid dimming molecules. The absolute interline background is 560\,$\pm$\,120\,photons\,s$^{-1}$\,m$^{-2}$\,$\micron^{-1}$\,arcsec$^{-2}$ under dark conditions. This value is similar to previous measurements without OH suppression suggesting that non-suppressed atmospheric emission is responsible for the interline background. Future OH suppression fibre designs may address this by the suppression of more sky lines using more accurate sky line measurements taken from high resolution spectra. 

\end{abstract}

\begin{keywords}
infrared: general -- instrumentation: miscellaneous -- instrumentation: spectrographs -- atmospheric effects 
\end{keywords}

\section{Introduction}

The near-infrared (NIR) background is $\approx$1000 times brighter than the optical background when observing from the ground, severely limiting our ability to observe faint objects at these wavelengths. For example, the average dark sky brightness in the $H$ band is 13.7\,mag\,arcsec$^{-2}$ (Vega system) compared to 21.8\,mag\,arcsec$^{-2}$ in the $V$ band \citep{sparke2007}. From NIR spectra of the night sky it is known that the background is dominated by the bright emission lines associated with the vibrational decay of atmospheric hydroxyl (OH) molecules \citep{meinel1950,dufay1951}. Complicating matters even further, the OH airglow is also known to fluctuate rapidly on short time-scales preventing clean subtraction from data \citep{davies2007}. The situation is rather unfortunate given the importance of the NIR window to the future of cosmology. Many key ultraviolet and optical spectral signatures used to study various astrophysical phenomena, notably star formation, are redshifted into the NIR window due to the expansion of the Universe. Thus, overcoming the OH airglow is a fundamental challenge for observational astronomy.



Space-based observations naturally provide a much fainter NIR background. Here the low background consists mainly of zodiacal scattered light and may be as low as $\approx$\,37.5 to 55\,photons\,s$^{-1}$\,m$^{-2}$\,$\micron^{-1}$\,arcsec$^{-2}$ in the $H$ band as measured from the Hubble Deep Field-North (ecliptic latitude $b=57$\,deg) and from the Hubble Ultra Deep Field ($b=-45$\,deg) using the F160W  filter ($\approx$\,0.4\,$\micron$ bandwidth) of NICMOS on the {\it Hubble Space Telescope} \citep{thompson2007}. 

In comparison, \citet{maihara1993} measure an interline sky brightness of 590\,$\pm$\,140\,photons\,s$^{-1}$\,m$^{-2}$\,$\micron^{-1}$\,arcsec$^{-2}$ at 1.665\,$\micron$ in the $H$ band from the ground using the University of Hawaii 2.2-metre telescope on Mauna Kea at a resolving power of $R=17,000$. More recently, \citet{ss2012} measure 670\,$\pm$\,200\,photons\,s$^{-1}$\,m$^{-2}$\,$\micron^{-1}$\,arcsec$^{-2}$ between 1.662 and 1.663\,$\micron$ using the Folded-port Infrared Echellette (FIRE) spectrometer at Magellan at $R=6000$. \citet{cuby2000} measure 1200\,photons\,s$^{-1}$\,m$^{-2}$\,$\micron^{-1}$\,arcsec$^{-2}$ in the $H$ band during the commissioning of the Infrared Spectrometer And Array Camera (ISSAC) on the ESO/VLT. Although the background in space is much darker when compared to the ground, the limited accessibility of space telescopes and the fact that the largest telescopes will always be on the ground drives astronomers to continue searching for ways to mitigate the OH airglow from the ground. 


Given that the OH lines are intrinsically narrow \citep{tl1983}, a low level background from the ground may be possible through OH suppression. It has been demonstrated that the Lorentzian wings of a diffraction grating line profile contribute non-negligible amounts of light from a bright emission line to the interline regions \citep{jbh2004,sce2008}. Recently, \citet{ss2012} claimed that the interline background in their measured $H$ band spectrum is consistent with just instrumentally-scattered OH emission. They compared their observed spectrum to a simulated sky spectrum generated from the OH line list of \citet{rousselot2000} with Gaussian line profiles ($R=30,000$) convolved with the measured FIRE line spread function. If this is the case then suppressing the OH lines before any dispersive element should limit the amount of OH light scattered by the diffraction grating into the interline regions providing higher sensitivity observations. For example, the simulations of \citet{sce2008} show that the background reduction when observing with pre-dispersion OH suppression together with adaptive optics (AO) versus observations with AO-only may be a factor of up to $\sim$\,7 greater than the background reduction when observing with AO versus natural seeing. Note that AO reduces the background because a smaller aperture may be used to sample the object compared to natural seeing. Thus, OH suppression is as important to cosmology as AO \citep{sce2008}. 

\citet{sce2012} and \citet{cqt2013} attempted to use OH suppression fibres with fibre Bragg gratings (FBGs) designed to suppress the 103 brightest OH doublets in the $H$ band (1.47--1.7\,$\micron$) before dispersion to reduce the Lorentzian wings and demonstrate an interline background reduction to produce higher sensitivity NIR observations. FBGs are the most advanced filter-based OH suppression technique to date \citep{jbh2011}. They do not have line profiles with scattering wings like diffraction gratings, which is the reason they provide an advantage \citep{sce2008}, in theory, over dispersion-based OH suppression techniques \citep[e.g.][]{iwamuro2001}. Here, the word grating refers to a refractive index modulation within the fibre core that induces Fresnel reflections. The filter response of an FBG is due to interference between the forward and backward propagating waves within the fibre core, much like a standard dielectric filter \citep{jbh2011}. The refractive index modulation ``grating''  of an FBG does not disperse light generating line profiles with scattering wing as a diffraction grating does. In this respect, they behave identically to ordinary optical fibres. 

$H$ band spectra of the night sky taken with OH suppression fibres show that the fibres do an excellent job of suppressing the OH lines compared to a non-suppressed control spectrum taken through an ordinary fibre \citep{sce2012,cqt2013}. However, the interline background in the OH suppressed spectrum and the non-suppressed spectrum is nearly identical \citep{sce2012}. This result suggests that the interline background may be dominated by contributions from unsuppressed OH lines and/or other atmospheric sources and/or astronomical sources. However, these night sky observations are detector noise-dominated in the interline regions, due to low system throughput, possibly obfuscating the expected interline background reduction. Plans are in place for these observations to be repeated in a regime that is sky background-limited using the same OH suppression fibres with an optimised fibre-fed spectrograph in early 2014 \citep{horton2012}. 

If the lack of a reduction in interline background is confirmed by these new observations, it may still be possible to achieve the desired background reduction and the associated increase in sensitivity by suppressing more lines if non-suppressed OH lines are responsible. If other atmospheric sources and/or astronomical sources are responsible, these sources may be filterable depending on whether they are line or continuum sources. Understanding the nature of the interline sky background in the OH suppressed observations made with the first generation of OH suppression fibres is an essential step toward perfecting this technology.

In this paper, we present the analysis of 19 hours of night sky observations in the $H$ band using the GNOSIS OH suppression unit with the IRIS2 imaging spectrograph \citep{tinney2004} at the 3.9-metre Anglo-Australian Telescope (AAT). In Section \ref{section:gnosis}, we give a brief description of GNOSIS followed by a description of our observations, data reduction and measurements in Section \ref{section:observations}. Section \ref{section:ILBsources} discusses the NIR background sources considered in this study, which includes non-thermal atmospheric emission, zodiacal scattered light and scattered moonlight. We discuss our results in Section \ref{section:results}, which includes an examination of the temporal behaviour of OH emission, a direct comparison of the spatial and temporal behaviours of the interline background with OH and O$_{2}$ emission, an assessment of the amount of zodiacal scattered light and scattered moonlight contributing to the interline background as well as a measure of the absolute interline sky brightness during dark time. A summary is given in Section \ref{section:summary}.

\begin{table*}
 \centering
 \begin{minipage}{140mm}
  \caption{Night sky observations made with GNOSIS+IRIS2 at the AAT. Symbol definitions are as follows: $t_{\mathrm{exp}}$ - exposure time, $t$ - time after sunset, $z$ - zenith distance, $l$ - longitude, $b$ - latitude, $\rho$ - lunar distance, $\alpha$ - lunar phase angle.\label{table:skytable}}
  \begin{tabular}{@{}cccccccccccc@{}}
  \hline
     &  &  &  & \multicolumn{5}{l}{Sky} & \multicolumn{3}{l}{Moon} \\
   &  &  &  & Local  &  \multicolumn{2}{l}{Galactic} & \multicolumn{2}{l}{Ecliptic} &  \\
   No.	& Date	& $t_{\mathrm{exp}}$ & $t$ & $z$  &$l$ & $b$ &  $l$ & $b$ & $\rho$ & $z$ & $\alpha$ \\
	&	& (s)	& (hr) & (deg) & (deg) & (deg) & (deg) & (deg) & (deg) & (deg) & (deg) \\
 \hline
 1 & 1 Sep 2011 & 1800 & 5.55 & 6.74 & 27.29 & -59.64 & 330.63 & -15.81 & 120.76 & 111.73 & -131 \\
 2 & 1 Sep 2011 & 1800 & 6.06 & 37.18 & 27.29 & -59.64 & 330.63 & -15.81 & 120.76 & 116.94 & -131 \\
 3 & 1 Sep 2011 & 1800 & 8.66 & 43.63 & 27.29 & -59.64 & 330.63 & -15.81 & 120.76 & 134.09 & -131 \\
 4 & 1 Sep 2011 & 1800 & 9.16 & 10.16 & 27.29 & -59.64 & 330.63 & -15.81 & 120.75 & 134.49 & -131 \\
 5 & 1 Sep 2011 & 1800 & 10.74 & 11.19 & 246.08 & -53.73 & 33.84 & -57.56 & 118.19 & 128.58 & -131 \\
 6 & 1 Sep 2011 & 1800 & 11.25 & 48.90 & 246.08 & -53.73 & 33.83 & -57.56 & 118.19 & 124.80 & -131 \\
 7 & 2 Sep 2011 & 1800 & 1.29 & 51.76 & 332.20 & 36.64 & 221.84 & -5.42 & 2.31 & 46.12 & -119 \\
 8 & 2 Sep 2011 & 1800 & 2.31 & 7.82 & 338.67 & 28.10 & 231.91 & -7.00 & 10.91 & 59.14 & -119 \\
 9 & 2 Sep 2011 & 1800 & 3.23 & 24.38 & 7.61 & -20.24 & 288.32 & -8.90 & 66.46 & 70.94 & -119 \\
 10 & 2 Sep 2011 & 1800 & 4.11 & 18.93 & 36.41 & -57.87 & 332.05 & -10.96 & 109.33 & 81.93 & -119 \\
 11 & 2 Sep 2011 & 1800 & 7.16 & 34.71 & 27.34 & -59.63 & 330.64 & -15.78 & 107.31 & 115.77 & -119 \\
 12 & 2 Sep 2011 & 1800 & 8.40 & 41.17 & 27.26 & -59.62 & 330.61 & -15.82 & 107.28 & 125.33 & -119 \\
 13 & 2 Sep 2011 & 1800 & 8.90 & 64.65 & 27.26 & -59.62 & 330.61 & -15.82 & 107.28 & 127.96 & -119 \\
 14 & 2 Sep 2011 & 1800 & 10.76 & 33.15 & 27.29 & -59.64 & 330.63 & -15.81 & 107.30 & 129.13 & -119 \\
 15 & 3 Sep 2011 & 1800 & 3.22 & 26.69 & 27.29 & -59.64 & 330.63 & -15.81 & 93.98 & 57.97 & -105 \\
 16 & 3 Sep 2011 & 1800 & 3.72 & 7.93 & 27.29 & -59.64 & 330.63 & -15.81 & 93.98 & 64.37 & -105 \\
 17 & 3 Sep 2011 & 1800 & 6.11 & 13.14 & 27.29 & -59.64 & 330.63 & -15.81 & 93.98 & 93.33 & -105 \\
 18 & 3 Sep 2011 & 1800 & 6.61 & 33.44 & 27.29 & -59.64 & 330.63 & -15.81 & 93.98 & 98.95 & -105 \\
 19 & 3 Sep 2011 & 900 & 8.67 & 32.84 & 154.71 & -57.52 & 27.68 & -10.06 & 149.40 & 118.47 & -105 \\
 20 & 3 Sep 2011 & 900 & 8.92 & 33.42 & 154.71 & -57.52 & 27.68 & -10.06 & 149.40 & 120.28 & -105 \\
 21 & 3 Sep 2011 & 900 & 9.69 & 34.39 & 154.71 & -57.52 & 27.68 & -10.06 & 149.40 & 124.69 & -105 \\
 22 & 3 Sep 2011 & 900 & 9.95 & 51.78 & 154.71 & -57.52 & 27.68 & -10.06 & 149.40 & 125.73 & -105 \\
 23 & 4 Sep 2011 & 900 & 4.37 & 49.57 & 90.63 & -48.78 & 356.16 & 11.24 & 106.22 & 60.16 & -92 \\
 24 & 4 Sep 2011 & 900 & 4.62 & 36.45 & 90.63 & -48.78 & 356.16 & 11.24 & 106.22 & 63.35 & -92 \\
 25 & 4 Sep 2011 & 900 & 5.78 & 35.19 & 88.21 & -55.56 & 355.40 & 4.34 & 105.63 & 77.71 & -92 \\
 26 & 4 Sep 2011 & 900 & 6.03 & 28.50 & 88.21 & -55.56 & 355.40 & 4.34 & 105.63 & 80.76 & -92 \\
 27 & 4 Sep 2011 & 1800 & 7.77 & 34.97 & 27.27 & -59.63 & 330.62 & -15.81 & 80.93 & 100.44 & -92 \\
 28 & 4 Sep 2011 & 1800 & 8.27 & 36.94 & 27.27 & -59.63 & 330.62 & -15.81 & 80.93 & 105.58 & -92 \\
 29 & 4 Sep 2011 & 900 & 10.33 & 38.76 & 154.71 & -57.52 & 27.68 & -10.06 & 136.69 & 121.83 & -92 \\
 30 & 4 Sep 2011 & 900 & 10.59 & 47.53 & 154.71 & -57.52 & 27.68 & -10.06 & 136.69 & 123.08 & -92 \\
 31 & 4 Sep 2011 & 900 & 11.54 & 4.28 & 154.71 & -57.52 & 27.68 & -10.06 & 136.69 & 125.78 & -92 \\
 32 & 5 Sep 2011 & 1800 & 0.48 & 49.45 & 353.78 & 5.13 & 259.21 & -8.29 & 9.54 & 9.04 & -80 \\
 33 & 5 Sep 2011 & 900 & 1.92 & 46.23 & 27.27 & -59.63 & 330.62 & -15.81 & 68.29 & 17.69 & -80 \\
 34 & 5 Sep 2011 & 900 & 2.17 & 31.68 & 27.27 & -59.63 & 330.62 & -15.81 & 68.29 & 20.72 & -80 \\
 35 & 5 Sep 2011 & 900 & 3.31 & 28.42 & 27.27 & -59.63 & 330.62 & -15.81 & 68.29 & 35.04 & -80 \\
 36 & 5 Sep 2011 & 900 & 3.56 & 18.59 & 27.27 & -59.63 & 330.62 & -15.81 & 68.29 & 38.29 & -80 \\
 37 & 5 Sep 2011 & 900 & 4.33 & 5.68 & 27.27 & -59.63 & 330.62 & -15.81 & 68.29 & 48.20 & -80 \\
 38 & 5 Sep 2011 & 900 & 5.62 & 6.12 & 27.27 & -59.63 & 330.62 & -15.81 & 68.29 & 64.55 & -80 \\
 39 & 5 Sep 2011 & 900 & 5.87 & 26.47 & 27.27 & -59.63 & 330.62 & -15.81 & 68.29 & 67.72 & -80 \\
 40 & 5 Sep 2011 & 900 & 7.65 & 29.72 & 27.29 & -59.63 & 330.63 & -15.81 & 68.30 & 89.16 & -80 \\
 41 & 5 Sep 2011 & 900 & 7.90 & 42.90 & 27.29 & -59.63 & 330.63 & -15.81 & 68.30 & 92.04 & -80 \\
 42 & 5 Sep 2011 & 900 & 8.93 & 46.13 & 27.29 & -59.63 & 330.63 & -15.81 & 68.30 & 103.10 & -80 \\
 43 & 5 Sep 2011 & 900 & 9.18 & 57.88 & 27.29 & -59.63 & 330.63 & -15.81 & 68.30 & 105.63 & -80 \\
 44 & 5 Sep 2011 & 900 & 10.11 & 61.05 & 27.29 & -59.63 & 330.63 & -15.81 & 68.30 & 114.06 & -80 \\
 45 & 5 Sep 2011 & 900 & 10.37 & 30.35 & 27.29 & -59.63 & 330.63 & -15.81 & 68.30 & 116.06 & -80\\
\hline
\end{tabular}
\end{minipage}
\end{table*}

\section{GNOSIS}\label{section:gnosis}
A detailed description of the GNOSIS OH suppression unit and its performance may be found in \citet{cqt2013} and \citet{sce2012}. Here we briefly describe the most important aspects. GNOSIS uses a fore-optics unit containing a 7-element hexagonal lenslet array integral field unit (IFU) mounted at the Cassegrain focus to collect light from the AAT. The hexagonal lenslets are hexagonally-packed and span 1.2\,arcsec (0.4\,arcsec for each lenslet flat-to-flat) on the sky. The light collected by the IFU is transported by a fibre bundle consisting of 50\,$\micron$ core diameter multi-mode fibres (MMFs) to the grating unit. 

The grating unit contains seven OH suppression fibres consisting of two photonic lanterns \citep{sls2005} and two aperiodic FBGs \citep{jbh2004}. The OH suppression fibres behave like a 50\,$\micron$ core diameter MMF with the spectral response of a single-mode fibre (SMF). The fibres suppress the 103 brightest OH doublets in the range 1.47--1.7\,$\micron$ using deep (up to 40\,dB depending on the line strength) and narrow ($R\approx$\,10,000) notches while maintaining relatively high throughput between the notches for scientific observations ($\approx$\,60 per cent at 1.55\,$\micron$). 

\begin{figure*}
\center
\begin{tabular}{c}
\includegraphics[width=0.95\textwidth]{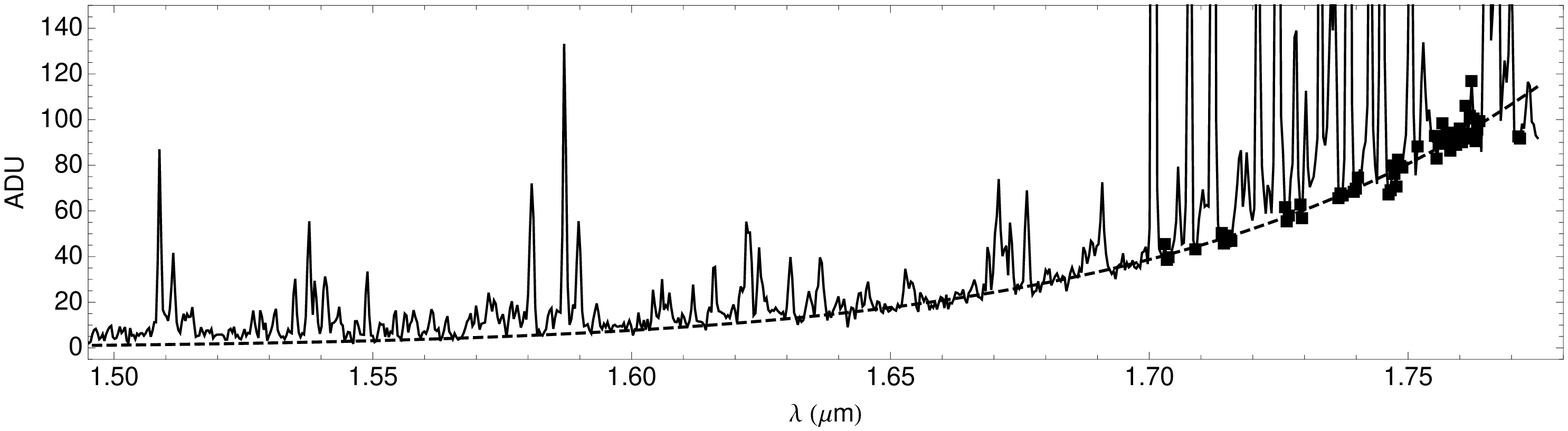} \\
(a)\\
\includegraphics[width=0.95\textwidth]{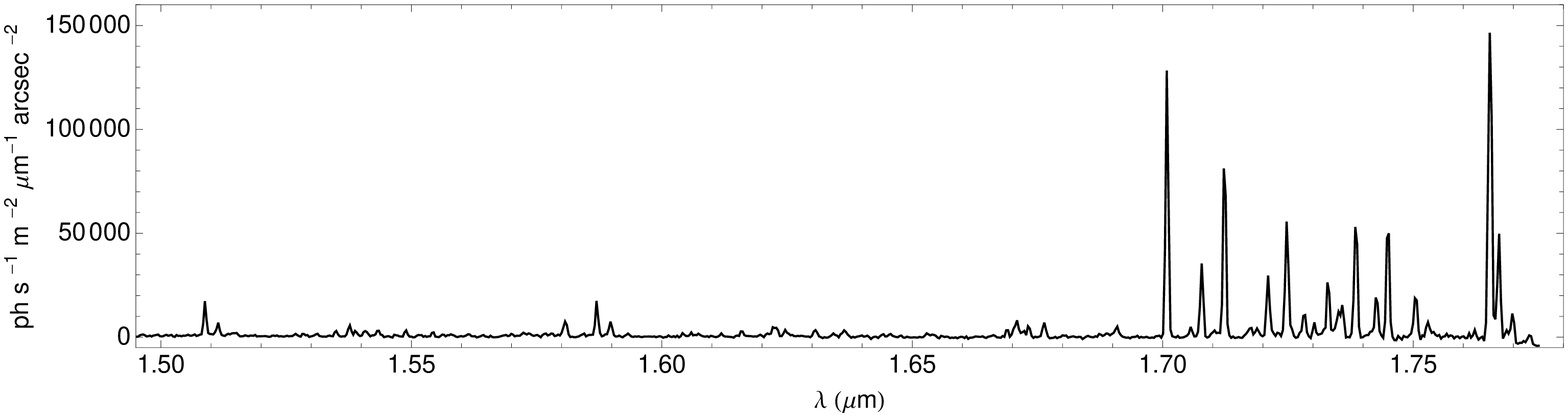} \\
(b)
\end{tabular}
\caption{(a) Dark-subtracted OH suppressed spectrum of sky frame 1 in Table \ref{table:skytable}. Squares show the continuum points used to fit a blackbody spectrum and the dashed line shows the best-fitting blackbody spectrum. (b) Dark-subtracted, thermal-subtracted and flux-calibrated OH suppressed spectrum of sky frame 1.\label{figure:skydark1}}
\end{figure*}

The grating unit passes the OH suppressed light to the IRIS2 spectrograph via a $\approx$\,10\,m long fibre bundle for measurement. The fibre bundle is terminated in a linear pseudo-slit mounted in an interface unit positioned over the IRIS2 dewar window. An optical relay within the interface unit images the pseudo-slit onto the IRIS2 slit plane with a PSF FWHM of $\approx$\,2\,pixels resulting in a resolving power of $R\approx$\,2350 and a dispersion of $\approx$\,3.5\,\AA\,pix$^{-1}$. IRIS2 contains a 1024$\times$1024 Rockwell Hawaii-1 detector with a dark current of $\approx$\,0.015\,$e^{-}$ s$^{-1}$ and an effective read noise of $\approx$\,8\,$e^{-}$ when using multiple read mode. For GNOSIS observations IRIS2 is configured with a custom slit mask with 250\,$\micron$ diameter holes and a $f$/12 cold stop. The total throughput of the system including atmosphere, telescope, GNOSIS and IRIS2 was $\approx$\,4 per cent for a diffuse source.

\section{Observations}\label{section:observations}
A total of 19 hours of $H$ band observations were taken between 1--5 September 2011 in multiple read mode (MRM) to minimise read noise. 45 separate sky frames were taken with exposure times of either 30 (61 reads, 30 s period) or 15 (61 reads, 15 s period) minutes. Table \ref{table:skytable} lists the details of each observation. The observations made on 1--4 September include six OH suppression fibres and one non-suppressed control fibre.

\subsection{Data Reduction}
We utilise a data reduction procedure similar to the general data reduction procedure for GNOSIS data given in \citet{sce2012}. Here we give a brief summary of the data reduction procedure specifically for the night sky observations. For each MRM sky frame all 61 reads are saved. There is a significant non-linearity in the detector response in the first few reads so we drop the first five reads from the linear least squares fit when constructing the final image. Then, we correct the image for the detector non-linearity due to the filling of the pixel wells by the procedure given on the IRIS2 web pages.\footnote{http://www.aao.gov.au/AAO/iris2/iris2\_linearity.html} 

The seven spectra are spread across two detector quadrants. We notice that despite using MRM, the median count level along the spectral direction computed from 200 spatial pixels at each spectral pixel in the un-illuminated portion of the detector shows a discontinuity in between the 2 quadrants and some slowly varying structure within each quadrant. We use this as our measure of dark current and subtract it from each detector row. 


	
The dark-subtracted sky spectrum in each fibre is extracted using the ``Gaussian summation extraction by least squares'' method of \citet{sb2010} using traces defined from a spectroscopic observation of the dome flat lamp. Spectroscopic observations of the dome flat lamp on each night are also used to determine the fibre-to-fibre throughput variation. The dome flat lamp spectrum in each fibre is extracted, integrated and normalised to the mean of all seven fibres. The extracted sky spectra are then divided by these relative throughput values. 

We obtain a wavelength calibration for each fibre from a xenon arc lamp observation. In the extracted arc spectrum of each fibre, we determine the pixel positions of the 1.4737, 1.5423, 1.6058 and 1.7330\,$\micron$ xenon lines by fitting a Gaussian to each line. The wavelength at each pixel $x$ is then given by the best-fitting cubic polynomial,
\begin{equation}
\lambda(x) = A+Bx+Cx^{2}+Dx^{3},
\end{equation} 
with $A$, $B$, $C$ and $D$ as unconstrained free parameters. 


The OH suppressed spectra are combined by taking the median of the six (1--4 September) or seven (5 September) spectra at each spectral pixel, which removes cosmic ray hits and other artefacts. The wavelength solution for the central sky fibre is applied to the median-combined OH suppressed spectrum. This results in an accuracy of $\approx$\,2\,$\mathrm{\AA}$ due to the slight shift in the wavelength solution from fibre to fibre.  

The GNOSIS+IRIS2 system is un-cooled, leading to a significant thermal background due to warm components. We remove the instrumental thermal background from each spectrum by fitting a blackbody spectrum with an emissivity of $\epsilon=1$ to the continuum values at $\lambda>1.7\,\micron$ using {\sc FindFit} in {\sc Mathematica} and subtracting the best-fitting model. The fitting procedure automatically accounts for the change in temperature throughout the night as each sky frame is treated separately. We choose the region beyond the OH suppression range where the thermal background is clearly dominant in order to avoid inadvertently removing sky emission in the suppression range where we wish to study the interline sky background. The functional form of the blackbody spectrum in units of photons\,s$^{-1}$\,m$^{-2}$\,$\micron^{-1}$\,arcsec$^{-2}$ is
\begin{eqnarray}
N_{\mathrm{bb}}(\lambda,T,\epsilon)&\approx&\frac{1.41\times10^{16}\epsilon}{\lambda^{4}\left[\exp\left(\frac{14387.7}{\lambda T}\right)-1\right]},
\end{eqnarray}
where $\lambda$ is the wavelength in $\micron$, and $T$ is the temperature. Most of the thermal background comes from the slit block in the IRIS2 interface unit and we convert the blackbody spectrum from photons\,s$^{-1}$\,m$^{-2}$\,$\micron^{-1}$\,arcsec$^{-2}$ to ADU assuming a throughput of 11.4 per cent, which includes a 5 per cent loss from slit block alignment and 12 per cent throughput for IRIS2 \citep{sce2012}. Figure \ref{figure:skydark1}a shows the OH suppressed spectrum from sky frame 1 in Table \ref{table:skytable} with the best-fitting blackbody model ($T\approx$\,284\,K) shown by the dashed curve and the continuum points used to compute the best-fitting model shown by squares.

Next, the sky spectra are divided by the instrument response, which corrects for variation in throughput with wavelength and telluric features. The instrument response is measured from a sky-subtracted A0V standard star observation. The stellar spectrum in each fibre is divided by a model spectrum of Vega \citep{ck1994} and normalised to the mean value between 1.5--1.69\,$\micron$. 

Lastly, the sky spectra are flux-calibrated assuming an efficiency of 4 per cent based on throughput measurements of GNOSIS, IRIS2, and estimates of the AAT \citep{sce2012,cqt2013}. Figure \ref{figure:skydark1}b shows the dark-subtracted, thermal-subtracted, flux-calibrated OH suppressed spectrum from sky frame 1. 

\begin{figure}
\includegraphics[width=0.47\textwidth]{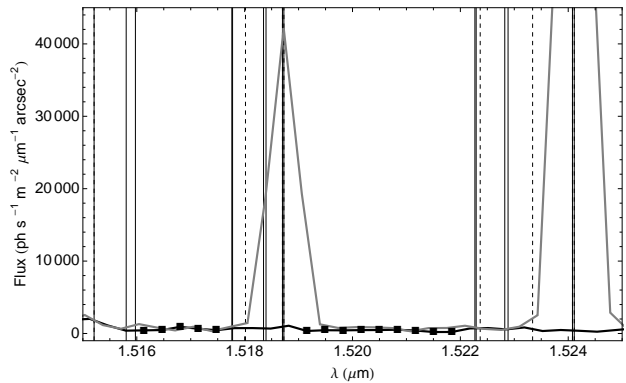} \\
\caption{OH suppressed (black) and non-suppressed (gray) spectra of sky frame 1 with the points used in the interline background measurement shown by squares. Vertical lines show the OH line positions from \citet{rousselot2000} and the dashed lines show the OH line positions from \citet{abrams1994}.\label{figure:sky1520}}
\end{figure}

\subsection{Interline background measurement}
We measure the interline background (ILB) in the OH suppressed spectrum of each sky frame. \citet{maihara1993} and \citet{ss2012} measure the interline background (ILB) at 1.665\,$\micron$ between two OH lines at 1.66111 and 1.66925\,$\micron$. In our data, this region is dominated by instrumental thermal background (see Figure \ref{figure:skydark1}a) making it difficult to assess the sky brightness at this wavelength due to uncertainties associated with the removal of the thermal background. First, from previous analysis of cold frames where the fore-optics unit is pointed directly at a container of liquid nitrogen we know that the thermal background for GNOSIS+IRIS2 observations is only approximately blackbody \citep[see][]{sce2012}. The discrepancy between the instrumental thermal background and a blackbody is significant in the Maihara region. Second, the blackbody fitting procedure uses the continuum values at $\lambda >$1.7\,$\micron$. As such, it removes the sky emission in this region. The procedure is necessary to provide a reasonable measure of the bright, non-suppressed OH lines at $\lambda >$1.7\,$\micron$, but we are over-subtracting the continuum in the Maihara region nearby. The instrumental thermal background measured from the cold frames are of limited use for thermal background removal. Very few cold frames were taken as they require removing the fore-optics unit from the AAT. Additionally, the cold frames were not taken simultaneously with the night sky observations and temperature changes result in significant discrepancies between the thermal background in each sky frame and the cold frames ($\approx$\,500\,photons\,s$^{-1}$\,m$^{-2}$\,$\micron^{-1}$\,arcsec$^{-2}$ for 2\,K).

To overcome the uncertainties associated with the removal of the thermal background, we utilise the blackbody fitting procedure described above as it automatically accounts for the change in thermal emission with temperature. Then, we select a region near 1.520\,$\micron$ where the thermal background is very low ($\approx$\,10\,photons\,s$^{-1}$\,m$^{-2}$\,$\micron^{-1}$\,arcsec$^{-2}$ at 282\,K). Specifically, we used the regions from 1.5160 to 1.5175 and 1.5190 to 1.5225\,$\micron$, which are near the suppressed 3--1 P2(1.5) and P1(2.5) OH lines at 1.51871 and 1.52410\,$\micron$, respectively, but do not coincide with these suppressed lines or any other non-suppressed OH lines that appear in the line lists of \citet{rousselot2000} or \citet{abrams1994}. The low thermal background in this region makes it less sensitive to the uncertainties associated with the thermal background removal. 

The OH suppressed (black) and non-suppressed control (gray) spectra of sky frame 1 are shown in Figure \ref{figure:sky1520} with the points used for the ILB measurement shown by squares. The OH lines according to \citet{rousselot2000} and \citet{abrams1994} are shown by the solid and dashed vertical lines, respectively. In the OH suppressed spectrum of each sky frame, we take the ILB to be the mean intensity in the two regions. The uncertainty in the ILB is given by the variation of this value over all six (1--4 September) or seven (5 September) OH suppressed fibres. Specifically, the uncertainty is the standard deviation of the mean intensity in the two regions in each fibre. 

\subsection{OH and O$_{2}$ measurement}
The non-suppressed $H$ band spectra of \citet{ss2012} appear consistent with an ILB dominated by instrumentally-scattered OH emission even at $R=6000$. The simulations of \citet{sce2008} show that even with OH suppression fibres, the ILB should still be dominated by residual OH emission. We measure several different sets of OH lines and O$_{2}$ in the OH suppressed spectrum of each sky frame and the non-suppressed control spectrum when available (1--4 September) to determine if our measured ILB contains instrumentally-scattered OH light and possibly other atmospheric emission from molecules with transitions near the ILB measurement region. The HITRAN2008 models \citep[High-resolution transmission molecular absorption database,][]{rothman2009} from SpectralCalc.com show 28, 106, 28, 30, 80, 15, and 46 transitions between 1.516 and 1.518\,$\micron$ for H$_{2}$O, CO$_{2}$, N$_{2}$O, CH$_{4}$, NO, OH and C$_{2}$H$_{2}$, respectively. Between 1.519 and 1.523\,$\micron$ the model shows 59, 234, 18, 48, 210, 43 and 148 transitions for those same species. Although the HITRAN2008 models include O$_{2}$, there are no transitions in the region of interest and O$_{2}$ is unlikely to directly contribute to the measured ILB. However, the ILB may contain contributions from any one of the molecules listed above, which may behave similar to O$_{2}$.


The two strongest OH lines near the ILB measurement region correspond to the 3--1 P2(1.5) and P1(2.5) transitions at 1.51871 and 1.52410\,$\micron$, respectively. Due to their proximity, these two lines are most likely to directly contribute instrumentally-scattered light to the measured ILB. We measure the peak value of these two lines in the control spectrum of the sky frames that included a control fibre. Note that these two lines cannot be measured in the OH suppressed spectrum because they are suppressed by the GNOSIS fibre design.


In the OH suppressed spectrum, the two strongest OH lines near the ILB measurement region correspond to the 3--1 Q1(3.5) and Q1(4.5) transitions at 1.50883 and 1.51137\,$\micron$, respectively. Although these two lines are less likely to directly contribute to the measured ILB because they are further away from the ILB measurement region, they provide a proxy for the 3--1 P2(1.5) and P1(2.5) lines on 5 September 2011. These lines are within the same vibrational transition and will be correlated. We measure the peak value of the 3--1 Q1(3.5) and Q1(4.5) lines in the OH suppressed spectrum of each sky frame. Note that these two lines are suppressed by the GNOSIS fibre design, but the suppression is poor due to the doublet separation being larger than the FBG notch width \citep{sce2012,cqt2013}.

According to the OH line list of \citet{abrams1994}, the lines near our ILB measurement region correspond exclusively to 2--0, 3--1 and 9--6 vibrational transitions. These transitions, especially the 3--1 vibrational transitions should exhibit a strong linear correlation with the measured ILB if these lines are directly contributing to the measured ILB compared to other vibrational transitions. To explore this, we measure the peak value of 17 very strong, non-suppressed OH lines ($\lambda >$1.7\,$\micron$) corresponding to 4--2, 5--3 and 6--4 vibrational transitions in the OH suppressed spectrum of sky frame. Table \ref{table:OHmeasurements} lists the wavelength of the peak of each measured OH lines and the corresponding quantum mechanical transitions for reference. 

We are mainly interested in the long-term variation of OH emission throughout the night. To reduce the rapid fluctuations of the individual OH lines we average lines of the same vibrational transition yielding three separate measures of OH emission. First, we average the 3--1 P2(1.5) and P1(2.5) lines to give what we refer to as the 3--1 P emission. Next, we average the 3--1 Q1(3.5) and Q1(4.5) lines to give what we refer to as the 3--1 Q emission. Lastly, we average the 17 lines at $\lambda >$1.7\,$\micron$ to give what we refer to as the 4--2, 5--3, 6--4 emission. 

Lastly, we measure the peak value of the a-X 1--0 O$_{2}$ line at 1.58\,$\micron$ in the OH suppressed spectrum of each sky frame. Note that this line is not suppressed in the GNOSIS fibre design \citep{sce2012}. 

\begin{table}
 \centering
 \begin{minipage}{80mm}
  \caption{Hydroxyl lines measured in GNOSIS data\label{table:OHmeasurements}}
  \begin{tabular}{@{}ccccc@{}}
  \hline
No. &   $\lambda$ & Transition & v$^{\prime\prime}$-v$^{\prime}$ & $J$\\
&    ($\micron$) & \\
 \hline
1 & 1.50883 & Q1 & 3--1 & 3.5 \\
2 &1.51137 & Q1 & 3--1 & 4.5 \\
3 & 1.51871 & P2 & 3--1 & 1.5 \\
4 & 1.52410 & P1 & 3--1 & 2.5 \\
5 & 1.70088 & P1 & 5--3 & 3.5 \\
6 & 1.70784 & P1, P2 & 4--2, 5--3 & 10.5, 3.5 \\
7 & 1.71237 & P1 & 5--3 & 4.5 \\
8 & 1.72103 & P2, R1 & 5--3, 6--4 & 4.5, 6.5 \\
9 & 1.72493 & P1 & 4--2, 5--3 & 11.5, 5.5 \\
10 & 1.72829 & R1 & 6--4 & 4.5 \\
11 & 1.73034 & R2 & 6--4 & 3.5 \\
12 & 1.73309 & R1 & 6--4 & 3.5 \\
13 & 1.73597 & R2 & 6--4 & 2.5 \\
14 & 1.73867 & P1, R1 & 5--3, 6--4 & 6.5, 2.5 \\
15 & 1.74270 & R2, P1 & 6--4, 4--2 & 1.5, 12.5 \\
16 & 1.74500 & R1 & 6--4 & 1.5 \\
17 & 1.75059 & P2 & 5--3 & 6.5 \\
18 & 1.75283 & P1 & 5--3 & 7.5 \\
19 & 1.76532 & Q2, Q1 & 6--4 & 0.5, 1.5 \\
20 & 1.76718 & Q1 & 6--4 & 2.5 \\
21 & 1.76984 & Q1 & 6--4 & 3.5 \\
\hline
\end{tabular}
\end{minipage}
\end{table}

\section{Interline background sources}\label{section:ILBsources}
Below we discuss the main NIR interline background sources, which include non-thermal atmospheric emission (OH and other molecules), zodiacal scattered light and scattered moonlight. Other contributions, such as zodiacal emission from interplanetary dust at the ecliptic plane, Galactic dust emission from interstellar dust and the cosmic microwave background are negligible in the $H$ band \citep{sce2008}. We removed thermal emission as fully as possible during our data reduction. We do not explicitly account for light pollution, which comes mainly from the town of Coonabarabran located $\approx$\,30\,km from Siding Spring Observatory. \citet{krisciunas2010} find measurable light pollution in the $V$ band from the city of La Serena located 55\,km away from Cerro Tololo Inter-American Observatory at zenith distances of 80\,deg or greater. In the $H$ band, the sources of light pollution (mostly visible) are substantially fainter and the Mie scattering efficiency by atmospheric aerosols decreases by a factor of $\approx$\,4. Hence, light pollution should only be substantial very close to horizon. Approximately 90 per cent of our $H$ band observations are taken at zenith distances smaller than 50\,deg and light pollution should be quite small in our sky frames. 

\subsection{Non-thermal atmospheric emission}\label{section:atmosphericmodel}

The NIR background is the $H$ band dominated by non-thermal atmospheric emission from OH and has contributions from other atmospheric molecules. Assuming a smooth distribution of these emitters, the emission intensity rises towards the horizon due to the airmass as the column density along the line of sight increases. The emission intensity also varies with time as a result of physical processes that affect that reactions rates between atmospheric molecules changing the number of emitters at any given time. As such, for a smooth distribution of emitters, the temporal and spatial parts may be considered independent and the emission from an atmospheric molecule may be written as
\begin{equation}\label{equation:atmosphericequation1}
I(t,z) = I_{0}(t)X(z),
\end{equation}
where $t$ is the time after sunset, $I_{0}(t)$ is the time-dependent intensity specific for each atmospheric molecule, $z$ is the zenith distance and $X(z)$ is the airmass. In reality, the spatial and temporal parts are not truly independent as OH emission has fine structure like cirrus clouds coving many degrees and this pattern drifts across the field of view over hours. However, \citet{ramsay1992} find no spatial variation of OH emission on a scale of 90\,arcsec in 80\,s exposures. Although our exposure times are longer (1800 or 3600\,s), our field of view is much smaller (1.2\,arcsec). Thus, the coupling of the spatial and temporal parts of OH emission due to the fine structure of the distribution of emitters is negligible in our observations.

For OH emission emitted from a layer $\approx$\,10\,km thick at height $h\approx$\,87\,km \citep{bs1988}, the airmass is given by the van Rhijn factor \citep{vanrhijn1921,ramsay1992,content1996} 
\begin{equation}\label{equation:x1}
X_{1}(z) = \left[1-\left(\frac{R}{R+h}\right)^{2}\sin^{2}z\right]^{-1/2},
\end{equation}
where $R$ is the radius of the Earth. For non-OH atmospheric emission, such as O$_{2}$ or in cases where we do not know the identity of the emitting species, we use the standard expression for airmass is given by 
\begin{equation}\label{equation:x2}
X_{2}(z)=\sec(z).
\end{equation}
However, note that in the range of zenith distances covered by our observations ($z\la$\,60\,deg), $X_{1}(z)$ and $X_{2}(z)$ are virtually identical differing by $\la$\,4 per cent. 

Excited OH radicals responsible for the OH airglow are primarily created by a reaction between hydrogen and ozone \citep{bates1950}
\begin{equation}
\mathrm{H}+\mathrm{O}_{3}\rightarrow \mathrm{OH}^{\ast} + \mathrm{O}_{2},
\end{equation}
and may be quenched by reactions such as \citep{llewellyn1978}
\begin{equation}
\mathrm{OH}^{\ast} + \mathrm{O}\rightarrow \mathrm{H} + \mathrm{O}_{2}.
\end{equation}
The temporal behaviour of OH emission varies on both short and long time-scales due to physical processes that affect these chemical reactions. On short time-scales, OH intensity fluctuates by $\approx$\,15 per cent over periods of 15 minutes \citep{taylor1991} to 1 hr \citep{yee1991} due to gravity waves \citep{ramsay1992,frey2000} that induce density and temperature perturbations. Note that the exposure times in our observations are roughly the same length as the period of these fluctuations, which serves to average out these variations as we are more interested in the long-term temporal behaviour of OH emission. 
 
On long time-scales, OH intensity decreases throughout the night by a factor of 2--3 \citep{ramsay1992,content1996,gbh2001} due to changes in the altitude-number density profiles of the minor atmospheric constituents as a result of the lack of sunlight \citep{sl1970}. The long-term decline of OH emission through the night is typically modelled with a simple monotonically declining linear model \citep{content1996,ss2012}.

\subsection{Zodiacal scattered light}\label{section:zslmodel}
\begin{figure}
\includegraphics[width=0.45\textwidth]{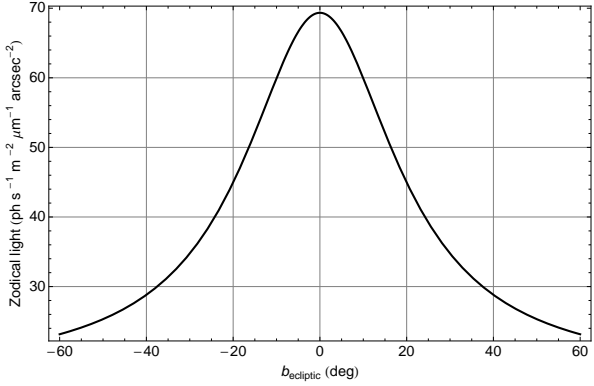}  \\
\caption{Zodiacal scattered light model from \citet{sce2008} based on $J$ band measurements from \citet{kelsall1998} with a $T=5800$\,K and $\epsilon=1.08\times10^{-13}$ blackbody spectrum.\label{figure:zslmodel}}
\end{figure}

Zodiacal scattered light (ZSL) from sunlight scattering off interplanetary dust is a large component of the NIR ILB in space-based observations. With OH suppression, ground-based observations may be able to reach background levels approaching ZSL levels. \citet{kelsall1998} provide a model for ZSL at 1.25\,$\micron$ as a function of ecliptic latitude ($b$) based on measurements from the {\it COBE} Diffuse Infrared Background Experiment (DIRBE). For the $H$ band, \citet{sce2008} utilise an analytic approximation to the \citet{kelsall1998} model with a $T=5800$\,K and $\epsilon=1.08\times10^{-13}$ blackbody spectrum, i.e. 
\begin{equation}\label{equation:zslmodel}
I_{\mathrm{ZSL}}(b)=N_{\mathrm{bb}}(\lambda,T,\epsilon)\left\{\frac{0.75}{[1+(2b/0.743)^{2}]}+0.25\right\}.
\end{equation}
Figure \ref{figure:zslmodel} shows the \citet{sce2008} ZSL model, which gives $\approx$\,70\,photons\,s$^{-1}$\,m$^{-2}$\,$\micron^{-1}$\,arcsec$^{-2}$ near the ecliptic plane. 

\subsection{Scattered moonlight}\label{section:moonlightmodel}
It is well-known that scattered moonlight, due to Rayleigh scattering by atmospheric gases and Mie scattering by atmospheric aerosols, contributes significantly to the sky background in the optical. Recently, \citet{ss2012} find correlations between lunar conditions and the ILB in the NIR. The correlations are moderate in the $Y$ band and decrease in strength at longer wavelengths due to the decrease in scattering efficiency with wavelength. We adapt the model of \citet{ks1991}, which is based on measurements in the $V$ band, to the $H$ band using the typical wavelength dependence of Rayleigh and Mie scattering as described in the Canada France Hawaii Telescope (CFHT) Redeye manual.\footnote{http://www.cfht.hawaii.edu/Instruments/Detectors/IR/\\Redeye/Manual/chapter7.html} The contribution from scattered moonlight varies as a function of wavelength ($\lambda$), lunar phase angle ($\alpha$), lunar zenith distance ($z_{\mathrm{moon}}$), the zenith distance of the sky position ($z$), lunar distance or the angular distance between the Moon and the sky position ($\rho$) and the atmospheric extinction ($k$) as
\begin{equation}\label{equation:bmoon}
B_{i}=I(\alpha)f_{i}(\rho,\lambda)10^{-0.4 k X_{3}(z_{\mathrm{moon}})}[1-10^{-0.4 k X_{3}(z)}],
\end{equation}
where the subscript $i=R$ refers to the contribution from Rayleigh scattering and $i=M$ and refers to the contribution from Mie scattering. $B_{i}$ is in the units of nanolamberts. 

The first term, $I(\alpha)$, is the intensity of moonlight as a function of phase angle in degrees and has the form
\begin{equation}
I(\alpha)=10^{-0.4[3.84+0.026|\alpha|+(4\times10^{-9})\alpha^{4}-1.3]}.
\end{equation}
The function is symmetric and sharply peaked about $\alpha=0$, which corresponds to full moon.

The second term, $f_{i}(\rho,\lambda)$, is the wavelength-dependent scattering function for either Rayleigh or Mie scattering. \citet{ks1991} measure $f_{R}(\rho)$ and $f_{M}(\rho)$ in the $V$ band over Mauna Kea. The wavelength dependence of Rayleigh scattering goes as $\lambda^{-4}$ and Mie scattering goes as $\lambda^{-1.3}$ \citep{cox2000}. Including these wavelength dependences normalised to 0.55\,$\micron$ in the $V$ band to the measured $f_{R}(\rho)$ and $f_{M}(\rho)$ of \citet{ks1991} yields
\begin{equation}
f_{M}(\rho,\lambda)=\left\{\begin{array}{ll}
6.2\cdot10^{7}\rho^{-2}\left(\frac{0.55}{\lambda}\right)^{1.3} & \rho<8.5\\
10^{6.15-\rho/40}\left(\frac{0.55}{\lambda}\right)^{1.3} & \rho\ge8.5 \\
\end{array}\right., 
\end{equation}
and
\begin{equation}
f_{R}(\rho,\lambda)=2.27\cdot10^{5}(1.06+\cos^{2}\rho)\left(\frac{0.55}{\lambda}\right)^{4},
\end{equation}
where $\lambda$ is in $\micron$ and $\rho$ is in degrees. 

The latter terms in Equation \ref{equation:bmoon} describe dependence of the scattered moonlight intensity with the zenith distance of the sky position and the lunar zenith distance, which arise due to airmass effects that attenuate the moonlight as it travels through the atmosphere before being scattered into the detector. \citet{ks1991} find that computing the airmass by
\begin{equation}\label{equation:x3}
X_{3}(z)=(1-0.96\sin^{2}z)^{-0.5},
\end{equation}
produces the best fit between the observed lunar brightness and the model brightness. 

\begin{figure}
\includegraphics[width=0.45\textwidth]{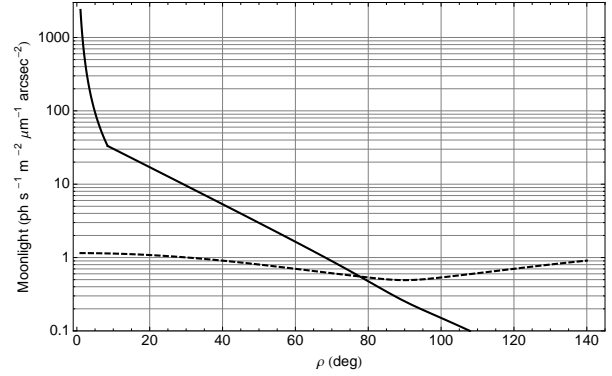} \\
\caption{Scattered moonlight model due to Rayleigh scattering by atmospheric gases (dashed) and Mie scattering by atmospheric aerosols (solid) for $\lambda=1.520\,\micron$, $\alpha=-80$\,deg, $z=0$\,deg, and $z_{\mathrm{moon}}=\rho$. Adapted from the $V$ band measurements of \citet{ks1991} using the typical wavelength dependence of Rayleigh and Mie scattering.\label{figure:moonlightmodel}}
\end{figure}

Equation \ref{equation:bmoon} may be converted to the units of mag\,arcsec$^{-2}$ by \citep{garstang1989}
\begin{equation}
V_{i}=\frac{20.7233-\ln(B_{i}/34.08)}{0.92104}.
\end{equation}
The final $H$ band scattered moonlight contribution is given in photons\,s$^{-1}$\,m$^{-2}$\,$\micron^{-1}$\,arcsec$^{-2}$ by
\begin{equation}\label{equation:ksmodel}
I_{\mathrm{moon}}^{i}(\lambda,\rho,\alpha,z_{\mathrm{moon}},z)=\frac{27.040\times10^{(20-V_{i})/2.5}}{0.290},
\end{equation}
Figure \ref{figure:moonlightmodel} shows $I_{\mathrm{moon}}^{R}$ and $I_{\mathrm{moon}}^{M}$ for $\lambda=1.520\,\micron$, $\alpha=-80$\,deg, $z=0$\,deg, $z_{\mathrm{moon}}=\rho$. $\alpha=-80$\,deg is the largest lunar phase angle in our data set and we only expect appreciable amounts of scattered moonlight at very small lunar distances due to Mie scattering by atmospheric aerosols.

\begin{figure}
\begin{tabular}{c}
\includegraphics[trim=0 36 0 0,clip,width=0.45\textwidth]{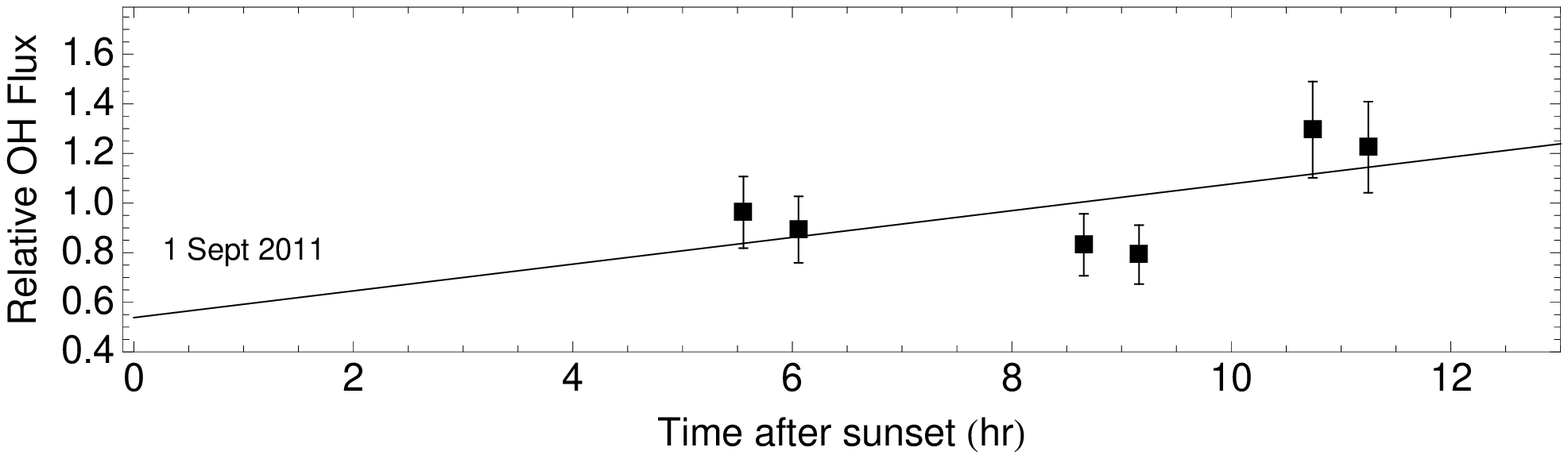} \\
\includegraphics[trim=0 36 0 0,clip,width=0.45\textwidth]{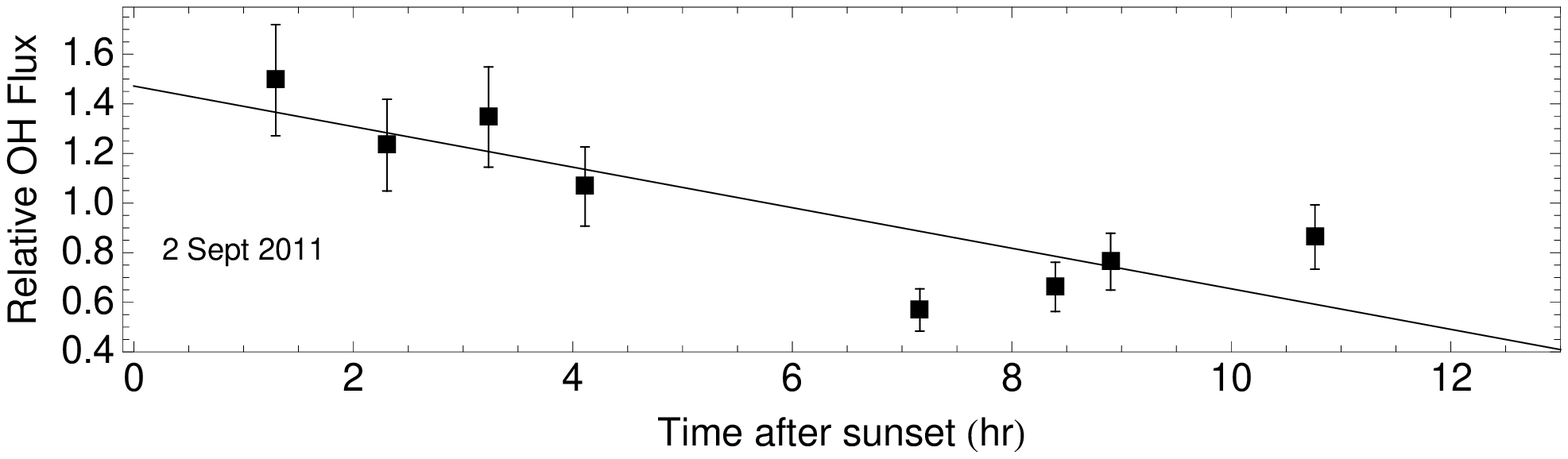}\\
\includegraphics[trim=0 36 0 0,clip,width=0.45\textwidth]{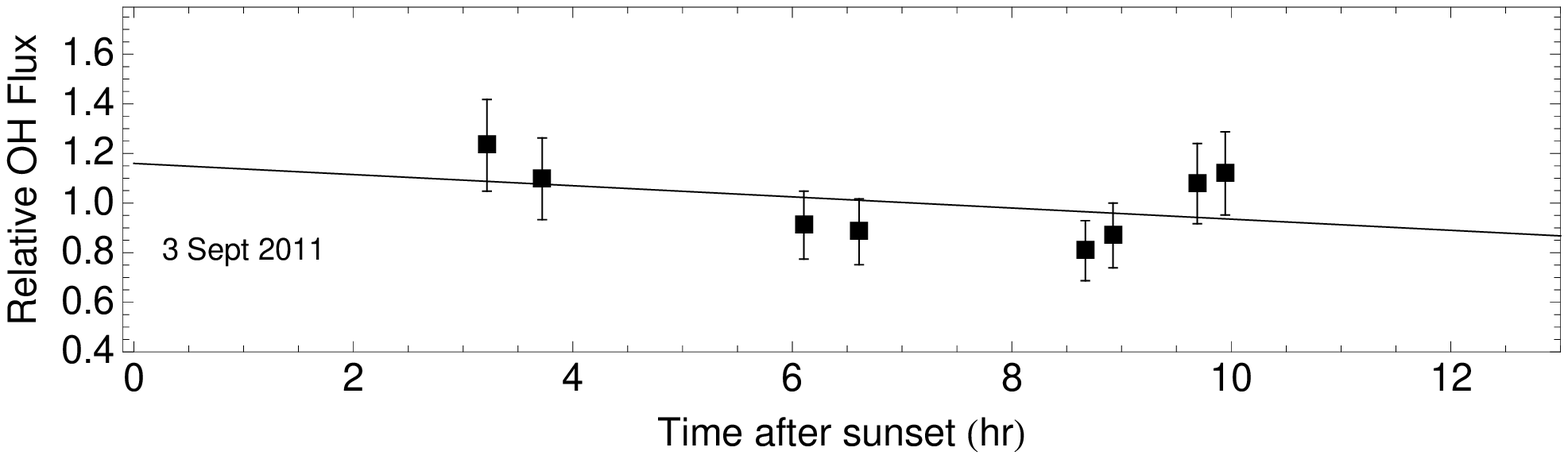}\\
\includegraphics[trim=0 36 0 0,clip,width=0.45\textwidth]{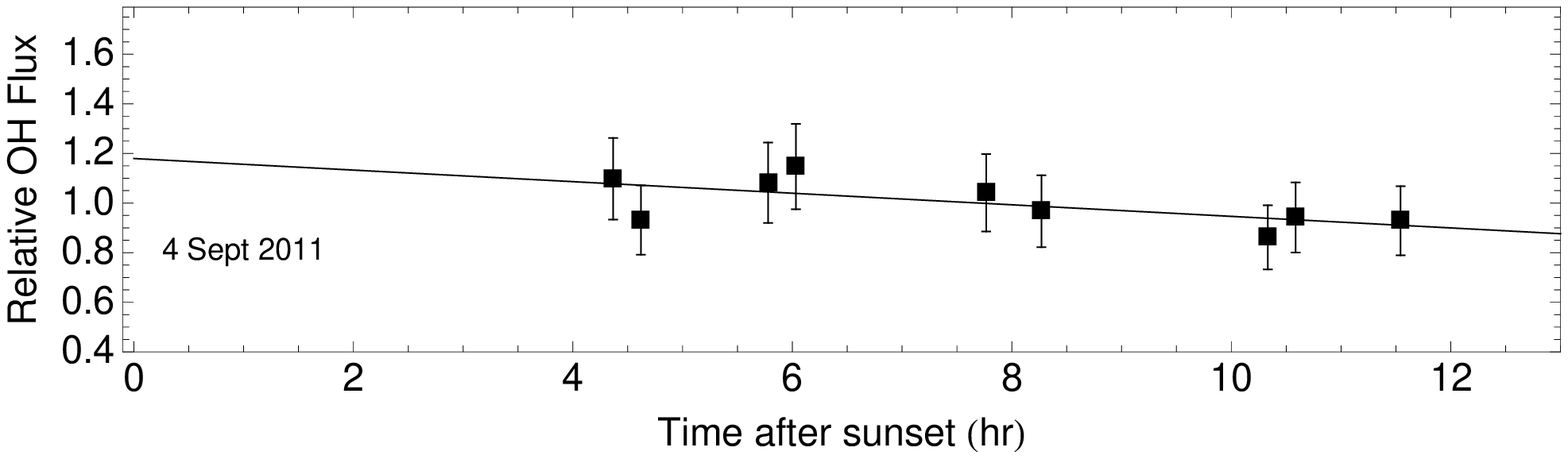}\\
\includegraphics[trim=0 0 0 0,clip,width=0.45\textwidth]{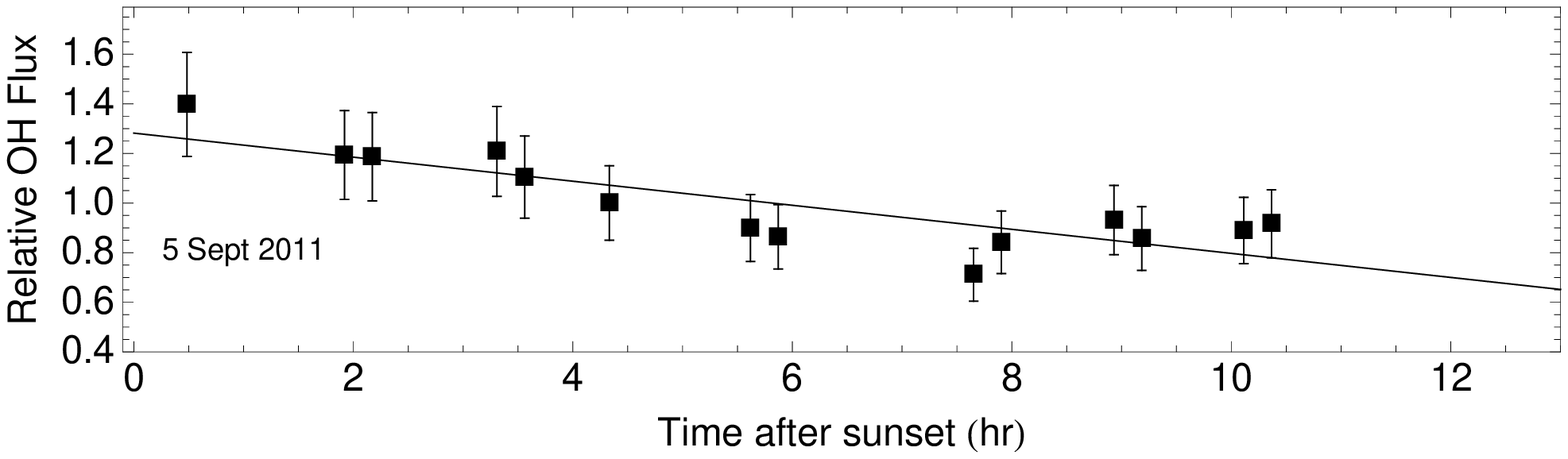}
\end{tabular}
\caption{3--1 Q OH emission for each night from 1 to 5 September 2011. Values are corrected to an airmass of unity by dividing by $X_{1}$ and normalised to the mean value on each night. Error bars represent the 15 per cent short time-scale fluctuations in OH emission due to gravity waves (lines are very bright and the Poisson noise is very small). Lines shows the best-fitting linear model. \label{figure:OHQ12timecurves}}
\end{figure}

\section{Results}\label{section:results}

\subsection{OH emission temporal behaviour}\label{section:OHtemporal}

First, we study the temporal behaviour of OH emission in order to obtain an empirical fit for $I_{0}(t)$, which will be used in Section \ref{section:ILBnonatmosphere} to separate the OH component of the ILB from the ZSL and scattered moonlight components. 


Figure \ref{figure:OHQ12timecurves} shows the temporal behaviour of the 3--1 Q OH emission for each night between 1 and 5 September 2011. The spatial dependence is removed by dividing by $X_{1}(z)$. Additionally, the intensities are normalised to the mean intensity on each individual night, which are listed in Table \ref{table:OHtable}. The temporal behaviour of OH emission is typically described by a simple linear model. We use {\sc LinearModelFit} in {\sc Mathematica} to determine the best-fitting linear model, shown by the line. The standard error for each free parameter is computed from the covariance matrix. The best-fitting slope and intercept, their standard errors and the coefficient of determination for each night are listed in Table \ref{table:OHtable}.  

As expected, the temporal behaviour of 3--1 Q OH emission is not simple. Although we do not show them here, the general temporal behaviours of 3--1 P and 4--2, 5--3, 6--4 OH emission are qualitatively very similar to the behaviour of 3--1 Q OH emission as shown in Figure \ref{figure:OHQ12timecurves}. 3--1 Q OH emission exhibits large intensity fluctuations despite our long integration times aimed at averaging out short time-scale fluctuations and the pattern is different from night to night. The time curves for 2 and 5 September 2011 show the 3--1 Q OH emission during the first half of the night being greater than the second half, in agreement with previous observations \citep{ramsay1992,content1996,gbh2001}. On the other hand, the time curves for 3 and 4 September 2011 are more level, but contain few sample points very early in the night when OH emission is the greatest. Our time curves are not regularly-sampled in time so we compute the RMS value to coarsely estimate the amplitude of the OH fluctuations over the entire night including both the long and short time-scale effects. The RMS values for each night vary from 9 to 32 per cent as listed in Table \ref{table:OHtable}. Despite its coarseness, the RMS values agree with the expected short time-scale variation of OH. For completeness, the RMS of the mean 3--1 Q OH emission on each night is 15 per cent and the total RMS over all nights is 27 per cent. These values are comparable to values of 20 and 30 per cent reported by \citet{content1996}.

\begin{figure}
\begin{tabular}{c}
\includegraphics[trim=0 36 0 0,clip,width=0.45\textwidth]{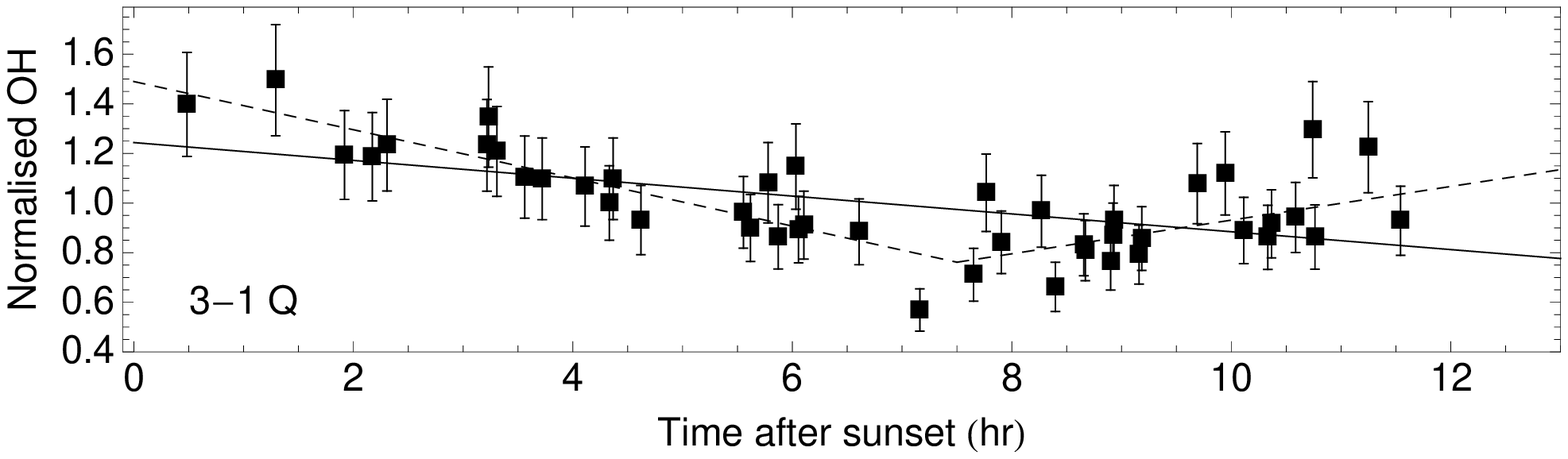} \\
\includegraphics[trim=0 36 0 0,clip,width=0.45\textwidth]{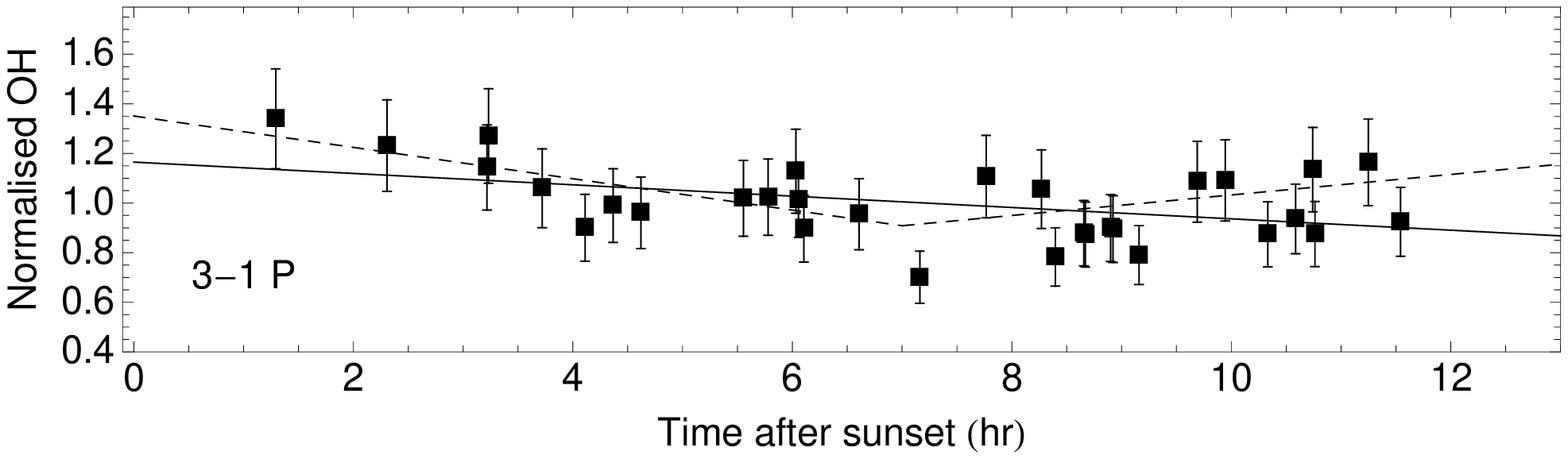}\\
\includegraphics[trim=0 36 0 0,clip,width=0.45\textwidth]{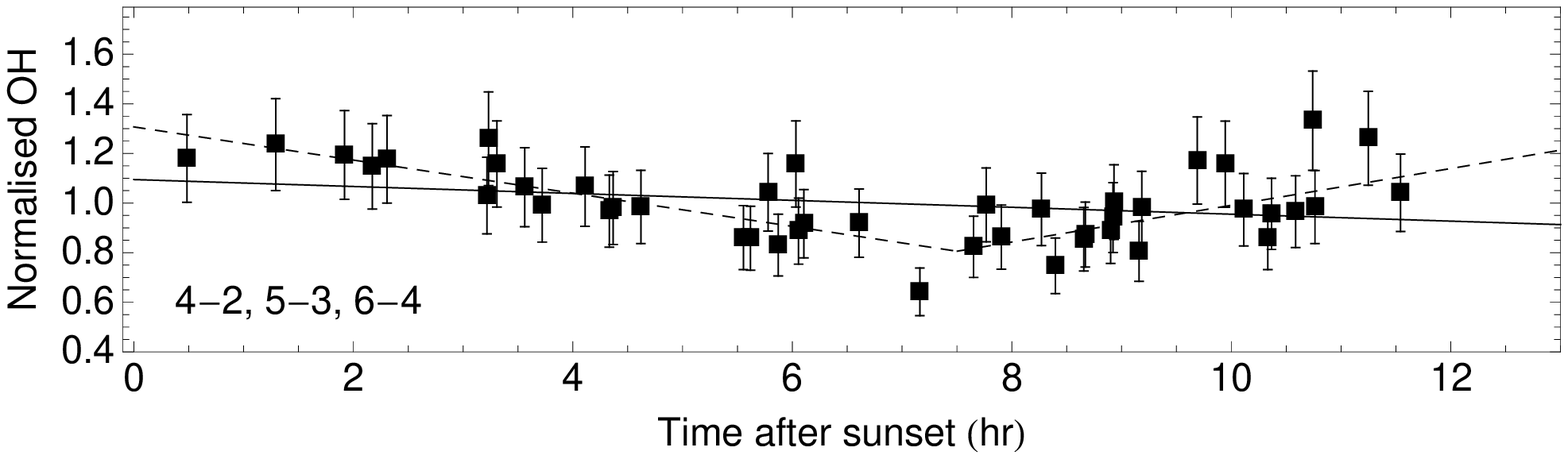}\\
\includegraphics[trim=0 0 0 0,clip,width=0.45\textwidth]{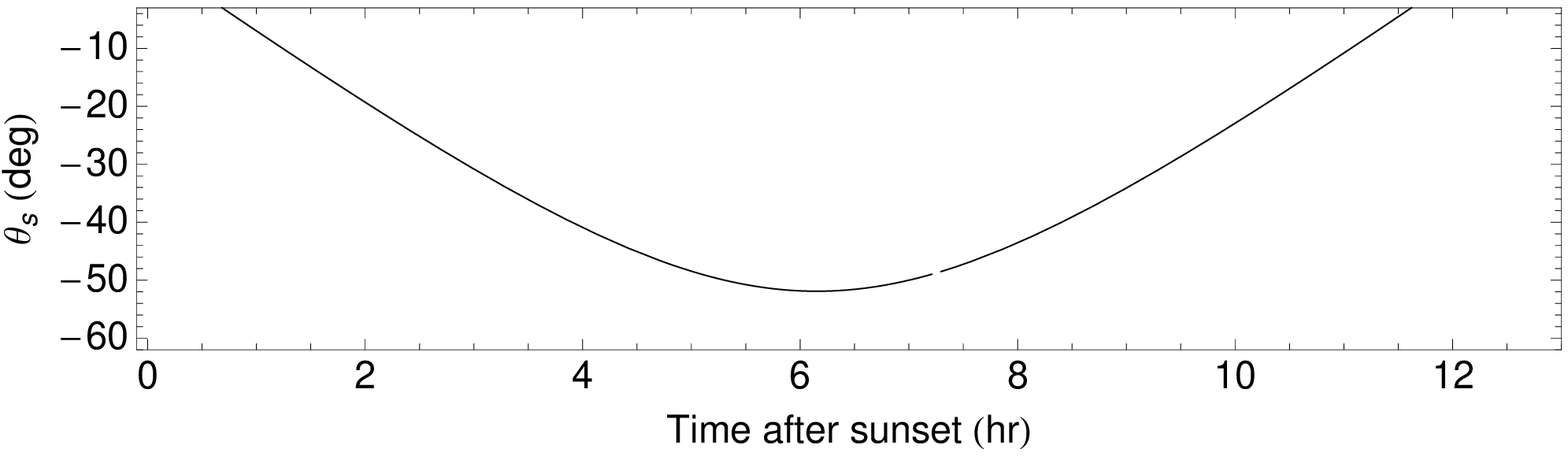}\\
\end{tabular}
\caption{Combined 3--1 Q (top), 3--1 P (second from top) and 4--2, 5--3 and 6--4 (third from top) OH emission from all available nights (1--5 September). Values have been corrected to an airmass of unity by dividing by $X_{1}$ and normalised to the mean value on each night for each set of OH lines. Error bars represent the 15 per cent short time-scale fluctuations in OH emission due to gravity waves. Line shows the best-fitting linear model. Bottom panel shows the solar elevation angle ($\theta_{s}$) on the night of 5 September. Dashed line shows a piecewise linear model where the first half of the night ($t\la$ 7.5\,hr) is treated separately from the second half and provides a better fit to the data compared to the simple linear model.\label{figure:OHAll}}
\end{figure}

To obtain an empirical fit for $I_{0}(t)$, we combine the data from all available nights such that the time curve is well-sampled. Figure \ref{figure:OHAll} shows the combined data for 3--1 Q, 3--1 P and 4--2, 5--3, 6--4 OH emission. The best-fitting linear model is shown by the line and the fit parameters are listed in Table \ref{table:OHcombined}. Note that the $R^{2}$ values for the simple monotonically decreasing linear model are relatively low for all three sets of OH lines.

\begin{table}
 \centering
 \begin{minipage}{120mm}
  \caption{Linear fit to 3--1 Q OH emission\label{table:OHtable}}
  \begin{tabular}{@{}ccccccc@{}}
  \hline
   Date & Mean$^{1}$ & RMS$^{2}$ & Slope  & Intercept & $R^{2}$ \\
 \hline
1 Sep 2011 & 12500 & 19 & 0.05\,$\pm$\,0.04 & 0.5\,$\pm$\,0.32 & 0.361 \\
2 Sep 2011 & 18700 & 32 & -0.08\,$\pm$\,0.02 & 1.5\,$\pm$\,0.14 & 0.708 \\
3 Sep 2011 & 15500 & 14 & -0.02\,$\pm$\,0.02 & 1.2\,$\pm$\,0.16 & 0.153 \\
4 Sep 2011 & 11600 & 9 & -0.02\,$\pm$\,0.01 & 1.2\,$\pm$\,0.08 & 0.427 \\
5 Sep 2011 & 14900 & 18 & -0.05\,$\pm$\,0.01 & 1.3\,$\pm$\,0.06 & 0.698 \\
\hline
\multicolumn{6}{l}{$^{1}$ photons\,s$^{-1}$\,m$^{-2}$\,$\micron^{-1}$\,arcsec$^{-2}$}\\
\multicolumn{6}{l}{$^{2}$ per cent}
\end{tabular}
\end{minipage}
\end{table}

Upon closer examination of the combined data for all three sets of OH lines, the data appears to decrease to a minimum near 7\,hr after sunset followed by subsequent rise in intensity. This behaviour is not adequately captured by the monotonically decreasing linear model as reflected by the low $R^{2}$ values. This rise in intensity later in the night may be a similar effect to the exponential rise in geocoronal H$\alpha$ due to solar Ly$\beta$ excitation of neutral hydrogen in the exosphere at a height of 500\,km \citep{jbh1998}. The increase in H$\alpha$ begins $\approx$\,6\,hr before sunrise as the amount of sunlight reaching the exosphere increases with increasing solar elevation angle ($\theta_{s}$). OH emission comes from an emitting layer in the mesosphere at $\approx$\,90\,km and it is plausible that it too increases as twilight ends due to increasing solar irradiation. To confirm this, we compute the solar elevation angle for the night of 5 September 2011, which is shown in bottom panel of Figure \ref{figure:OHAll}. Comparing the combined OH emission data with the solar elevation, the intensity is minimum near the minimum solar elevation angle at $\approx$\,7\,hr after sunset and the intensity rises after this time coinciding with an increase in solar elevation angle. Thus, a function that follows the behaviour of $\theta_{s}(t)$ is probably best suited to describing the OH emission intensity as a function of time. 

Fitting linear models separately to the first and second halves of the night gives a function with this behaviour. The best-fitting slope and intercept of this piecewise linear model are listed in Table \ref{table:OHcombined} and shown by the dashed line in Figure \ref{figure:OHAll}. The intercept of the linear model for the second half of the night is constrained to be continuous with the linear model for the first half of the night at either 7 or 7.5\,hr (whichever provides the best fit to the data). Simply comparing the $R^{2}$ values demonstrates that the piecewise linear model gives a better description of the temporal behaviour of OH emission than the simple monotonically declining linear model. We will use the best-fitting piecewise linear model for the 3--1 Q OH emission as our $I_{0}(t)$ in Section \ref{section:ILBnonatmosphere}.

\subsection{Interline background airmass dependence}\label{section:ILBspatial}
Now we determine the nature of the ILB when using OH suppression fibres. First, we directly compare the airmass dependence of the ILB, OH and O$_{2}$ emission to demonstrate that the ILB contains a strong atmospheric component. Both OH and O$_{2}$ exhibit a similar airmass dependences as both are atmospheric emission sources (but have different temporal behaviour early in the night).

We simplify Equation (\ref{equation:atmosphericequation1}) to a single variable by only considering 13 sky frames of the same field ($\mathrm{RA}=338.839$\,deg, $\mathrm{dec}=-25.953$\,deg), which we will refer to as field A from here on, taken throughout the night of 5 September 2011 as field A rises and sets. In this case, the zenith distance is a simple function of time after sunset given by
\begin{equation}
\cos z(t) = \sin\delta_{A}\delta_{\mathrm{sso}}+\cos\delta_{A}\cos\delta_{\mathrm{sso}}\cos[\alpha_{A}-\alpha_{\mathrm{sso}}(t)],
\end{equation}
where $(\alpha_{A},\delta_{A})$ and $(\alpha_{\mathrm{sso}}(t),\delta_{\mathrm{sso}})$ are the right ascension and declination of the field A and zenith at Siding Spring Observatory at the time of observation, respectively, and $\alpha_{\mathrm{sso}}(t)$ is the sidereal time. Hence, the emission intensity is only a function of time after sunset, i.e.,
\begin{equation}\label{equation:atmosphericemission}
I(t) = I_{0}(t)X[z(t)].
\end{equation}
A quick examination of Equations (\ref{equation:x1}), (\ref{equation:x2}) and (\ref{equation:x3}) shows that regardless of the specific expression used for the airmass, $X[z(t)]$ monotonically decreases to a minimum as field A rises ($z(t)$ decreasing) during the first half of the night and $X[z(t)]$ monotonically increases as field A sets ($z(t)$ increasing) during the second half of the night. This behaviour readily distinguishes an atmospheric source from non-atmospheric sources and we examine if the ILB follows this sort of curve to establish whether or not it is atmospheric in nature.

\begin{table}
 \centering
 \begin{minipage}{80mm}
  \caption{Fit parameters for combined OH emission data\label{table:OHcombined}}
  \begin{tabular}{@{}lccc@{}}
  \hline
   Transition & Slope & Intercept & $R^{2}$ \\
 \hline
3--1 Q  & -0.04\,$\pm$\,0.01 & 1.2\,$\pm$\,0.06 & 0.305 \\
3--1 Q ($<$ 7.5\,hr) &  -0.10\,$\pm$\,0.01 & 1.49\,$\pm$\,0.07 & 0.744 \\
3--1 Q ($>$ 7.5\,hr) & 0.07\,$\pm$\,0.03 & 0.25 & 0.252 \\
3--1 P  & -0.04\,$\pm$\,0.01 & 1.2\,$\pm$\,0.06 & 0.197 \\
3--1 P ($<$ 7\,hr) & -0.06\,$\pm$\,0.02 & 1.35\,$\pm$\,0.07 & 0.546 \\
3--1 P ($>$ 7\,hr) & 0.04\,$\pm$\,0.03 & 0.62 & 0.148 \\
4--2, 5--3, 6--4  & -0.04\,$\pm$\,0.01 & 1.2\,$\pm$\,0.06 & 0.080 \\
4--2, 5--3, 6--4 ($<$ 7.5\,hr) &  -0.07\,$\pm$\,0.01 & 1.31\,$\pm$\,0.05 & 0.627 \\
4--2, 5--3, 6--4 ($>$ 7.5\,hr) & 0.07\,$\pm$\,0.02 & 0.25 & 0.340 \\
\hline
\end{tabular}
\end{minipage}
\end{table}

\begin{figure}
\begin{tabular}{c}
\includegraphics[trim=0 36 0 0,clip,width=0.45\textwidth]{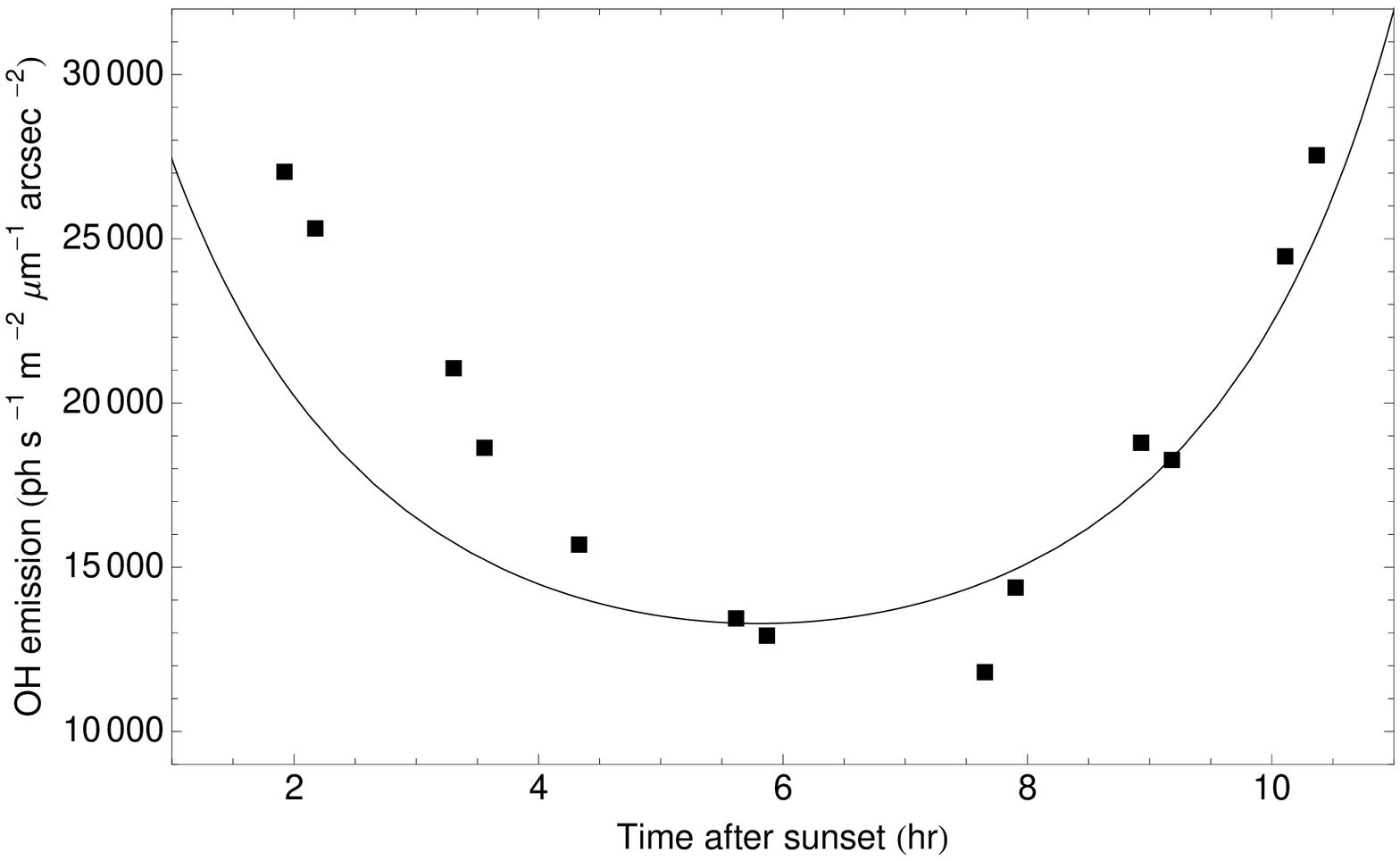} \\
\includegraphics[trim=0 36 0 0,clip,width=0.45\textwidth]{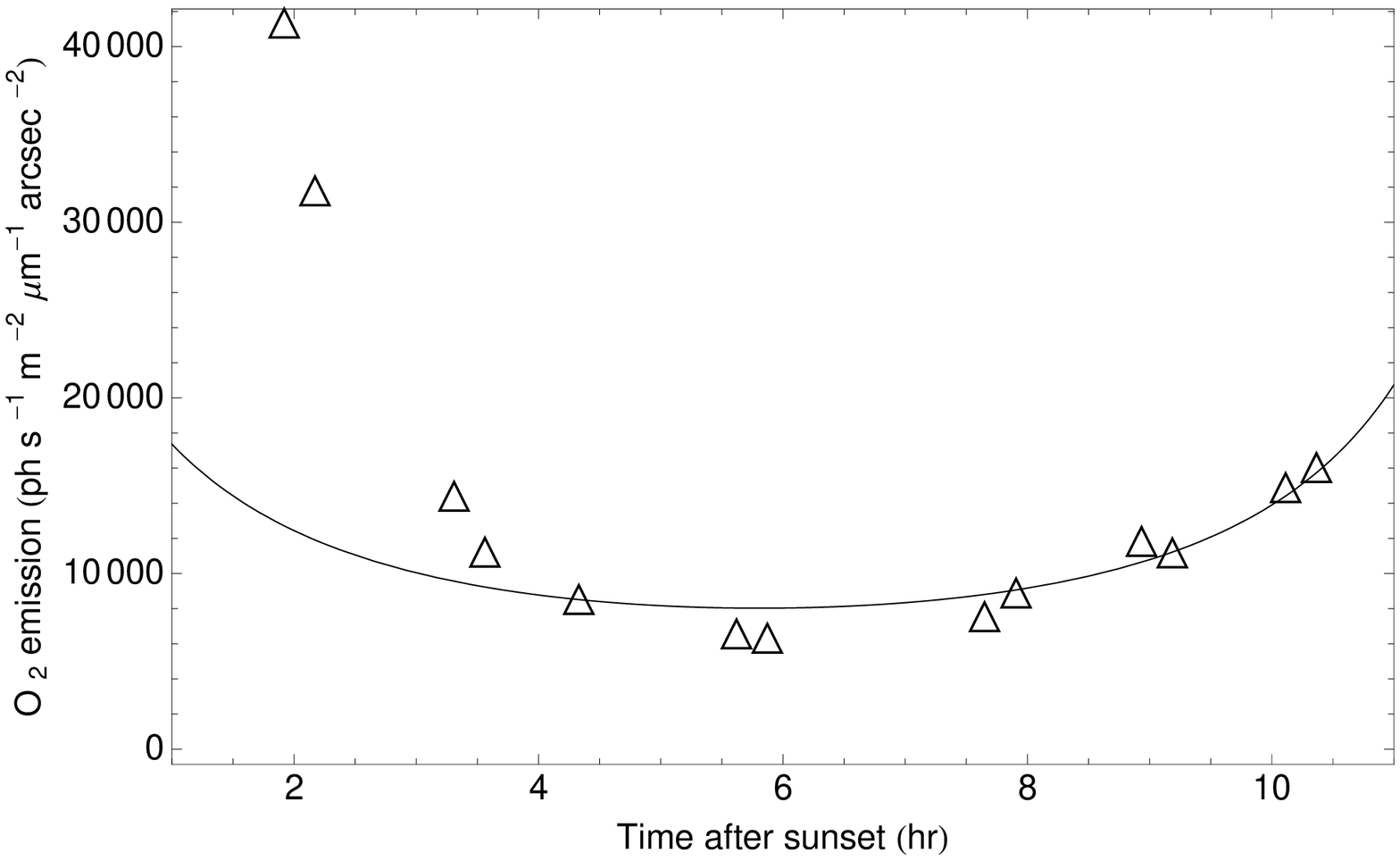} \\
\includegraphics[width=0.45\textwidth]{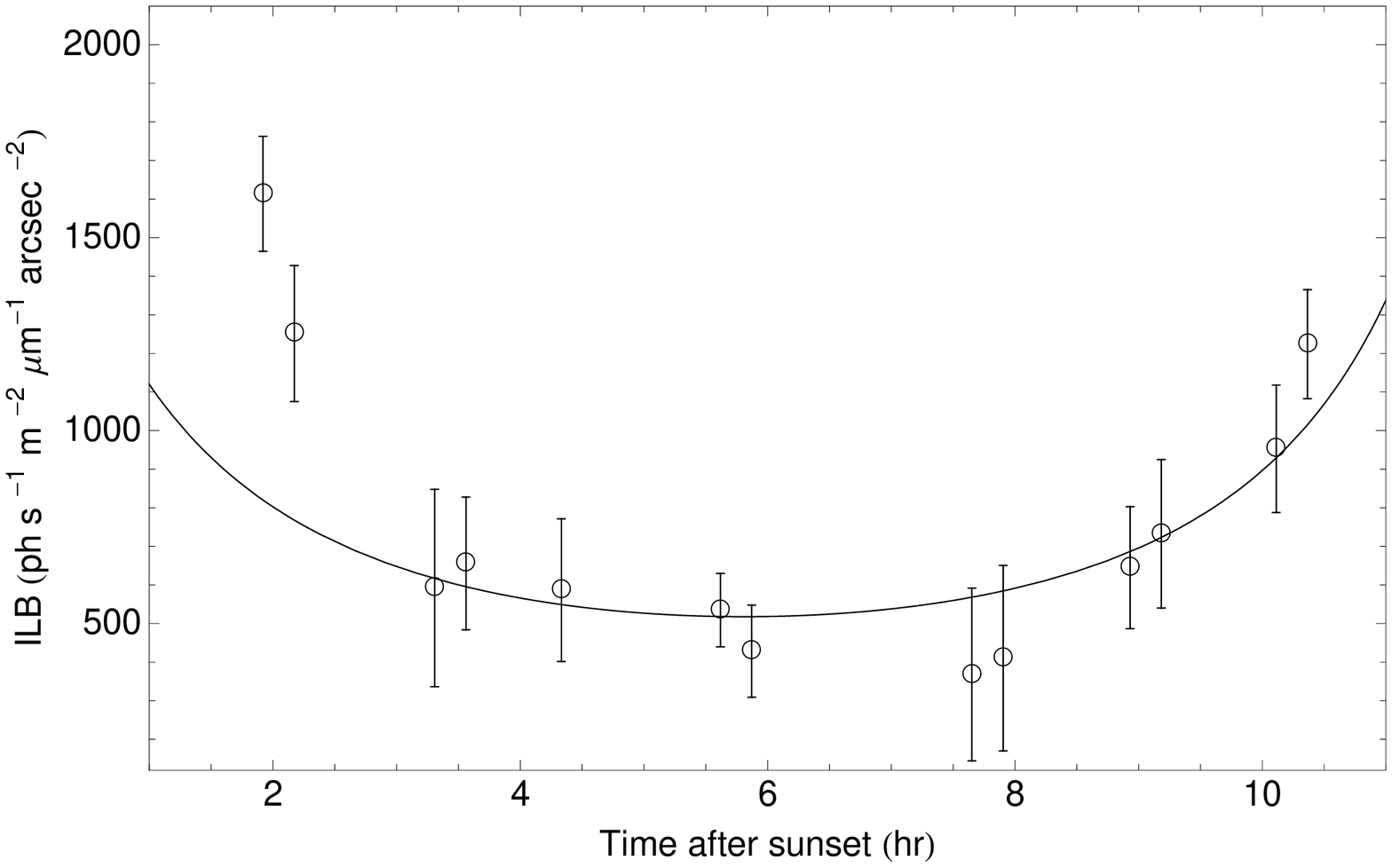} \\
\end{tabular}
\caption{3--1 Q OH (top) and O$_{2}$ (middle) emission and the interline background (bottom) as a function of time after sunset from 13 observations taken on 5 September 2011 following field A as it rises and sets. Line shows the $I(t)$ curve for an atmospheric source when following a rising and setting field with constant $I_{0}(t)$ (determined by eye) and with $X=X_{1}$ for OH and $X=X_{2}$ for O$_{2}$ and the interline background. OH, O$_{2}$ and the interline background are not truly constant $I_{0}(t)$ sources, which explains the deviations from the $I(t)$ curves. Regardless, the characteristic $X[z(t)]$ behaviour for an atmospheric source when following a rising and setting field is clearly visible. Note that OH and O$_{2}$ may fluctuate by $\approx$\,15 per cent due to gravity waves.\label{figure:timecurvesep05}}
\end{figure}

Figure \ref{figure:timecurvesep05} shows the 3--1 Q OH (top) and O$_{2}$ (middle) emission and the ILB (bottom) as a function of time after sunset. 3--1 P OH emission is not shown because the sky frames on 5 September 2011 did not include a control fibre. 4--2, 5--3, 6--4 OH emission is not shown but its behaviour is qualitatively very similar to that of 3--1 Q emission shown in Figure \ref{figure:timecurvesep05}. The lines show the $I(t)$ curve for an atmospheric source when following a rising and setting field with constant $I_{0}(t)$ and with $X=X_{1}$ for OH and $X=X_{2}$ for O$_{2}$ and the ILB. Recall that $X_{1}$ and $X_{2}$ are virtually identical over our range of zenith distances and our choice of $X_{2}$ for O$_{2}$ and the ILB does not affect our results or conclusions. The $I_{0}(t)$ values have been determined by eye. Both 3--1 Q OH and O$_{2}$ exhibit the expected $X[z(t)]$ behaviour, monotonically decreasing to a minimum and monotonically increasing during the second half of the night. The data do not exactly follow the curve, but this is expected as neither OH or O$_{2}$ are truly constant $I_{0}(t)$ sources. OH shows elevated emission both early and late in the night due to mechanisms discussed in Section \ref{section:atmosphericmodel} and O$_{2}$ shows very rapid dimming after sunset as previously observed by one of us (AJH) during the first on-sky observations with FBGs \citep{jbh2011}.

The ILB behaviour with time after sunset exhibits the $X[z(t)]$ behaviour expected for an atmospheric emission source like 3--1 Q OH and O$_{2}$ emission. This suggests that the ILB has a strong atmospheric component that varies according to $X[z(t)]$ (either $X_{1}$ or $X_{2}$), but with elevated emission early in the night due to an $I_{0}(t)$ term for the specific molecule(s) responsible for the ILB. The ILB trend with time after sunset is not consistent with ZSL, which is constant with time after sunset given that the ecliptic latitude for field A is fixed in this set of observations. Scattered moonlight varies with $X_{3}[z(t)]$ and $X_{3}[z_{\mathrm{moon}}(t)]$, but the large lunar phase angle ($-80$\,deg) and lunar distance (68\,deg) of these observations results in this component being much too weak (see Figure \ref{figure:moonlightmodel}) to account for the observed ILB intensity variation with time. Thus, from the airmass dependence of the ILB, we conclude that it is certainly dominated by emission from atmospheric molecules. 

\begin{figure}
\includegraphics[trim=5 0 0 0,clip,width=0.47\textwidth]{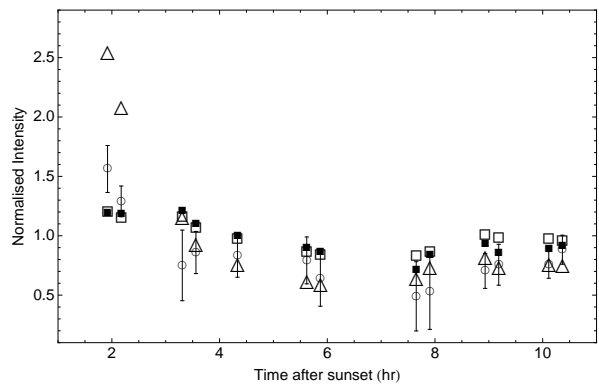} \\
\caption{Airmass-corrected 3--1 Q OH (filled squares), 4--2, 5--3, 6--4 OH (open squares) and O$_{2}$ (triangles) emission and the interline background (cricles) as a function of time after sunset normalised to the mean intensity for the 13 observations of field A on the night of 5 September 2011. Late in the night, the temporal behaviour of the interline background is similar to both OH and O$_{2}$. Early in the night, the dimming of the interline background is intermediate between the rapid rate of O$_{2}$ and the slow rate of OH suggesting contributions from multiple atmospheric molecules with different dimming rates early in the night. Note that OH and O$_{2}$ may fluctuate by $\approx$\,15 per cent due to gravity wave.\label{figure:timecurvesep05Xcorr}}
\end{figure}

\subsection{Interline background temporal behaviour}\label{section:ILBtemporal}
Having established that the dominant component of the ILB is atmospheric in origin, we search for evidence regarding which atmospheric molecule(s) are responsible for the ILB. To do this, we study the temporal behaviour, $I_{0}(t)$, of the ILB and make direct comparisons with 3--1 Q OH, 4--2, 5--3, 6--4 OH and O$_{2}$ emission. 

We obtain $I_{0}(t)$ for the ILB, OH and O$_{2}$ by dividing the intensity values by the airmass. OH is corrected using $X_{1}$ while O$_{2}$ and the ILB are corrected using $X_{2}$. The ILB could have been corrected using $X_{1}$ considering OH is the most logical atmospheric contributor. However, we avoid assuming that the dominant atmospheric component of the ILB is OH by using the standard airmass expression. In any case, as noted above, $X_{1}$ and $X_{2}$ are virtually identical over the range of zenith distances covered by our data and our choice of $X_{2}$ does not affect any of our results or conclusions. Figure \ref{figure:timecurvesep05Xcorr} shows the airmass-corrected ILB (circles), 3--1 Q OH (filled squares), 4--2, 5--3, 6--4 OH (open squares) and O$_{2}$ (triangles) emission intensities normalised to the mean intensity for the each night as a function of time after sunset mentioned earlier.


3--1 Q OH, 4--2, 5--3, 6--4 OH and O$_{2}$ emission all display the expected long-term decline in their emission intensities throughout the night due changes in the rates of the reactions producing these emitters from a lack of sunlight. The two sets of OH lines are nearly identical during the first half of the night, but show subtle differences in their temporal behaviour during the second half of the night. O$_{2}$ behaves very similarly to OH from $t\approx$\,3.5\,hr after sunset onwards. The most noticeable difference between OH and O$_{2}$ is the very rapid decline of O$_{2}$ emission immediately after sunset. 

The airmass-corrected ILB shows a clear dependence on time. In their moonless, non-suppressed FIRE spectra, \citet{ss2012} find a moderate linear correlation between the $H$ band continuum and local time ($R^{2}=0.461$). The best-fitting linear model to our measured ILB data shown in Figure \ref{figure:timecurvesep05Xcorr} yields $R^{2}=0.368$ in good agreement with \citet{ss2012}. However, the linear correlation of the ILB with time is of limited use in identifying the atmospheric emitter(s) responsible for the ILB, which is the main goal of this paper. 

Directly comparing the ILB to OH, we see that the two closely follow one another from $t\ga$4\,hr after sunset, suggesting that the ILB may indeed contain OH. However, the ILB is also consistent with O$_{2}$ in this range suggesting a possible contribution from an O$_{2}$-like molecule. Early in the night ($t\la4$\,hr after sunset), the behaviour of the ILB, OH and O$_{2}$ are very distinct providing valuable information on the composition of the ILB. OH is the most likely atmospheric contributor to the ILB, but the ILB dims more rapidly than OH early in the night. An ILB with contributions from slow dimming molecules such as OH and rapid dimming O$_{2}$-like molecules could explain the the intermediate dimming rate of the ILB seen early in the night in Figure \ref{figure:timecurvesep05Xcorr}. We conclude that the measured ILB has a definite time dependence, showing a decline during the first half of the night and rise during the second half of the night, which is consistent with the behaviour of atmospheric OH.
 

\begin{figure*}
\begin{tabular}{cc}
\includegraphics[width=0.47\textwidth]{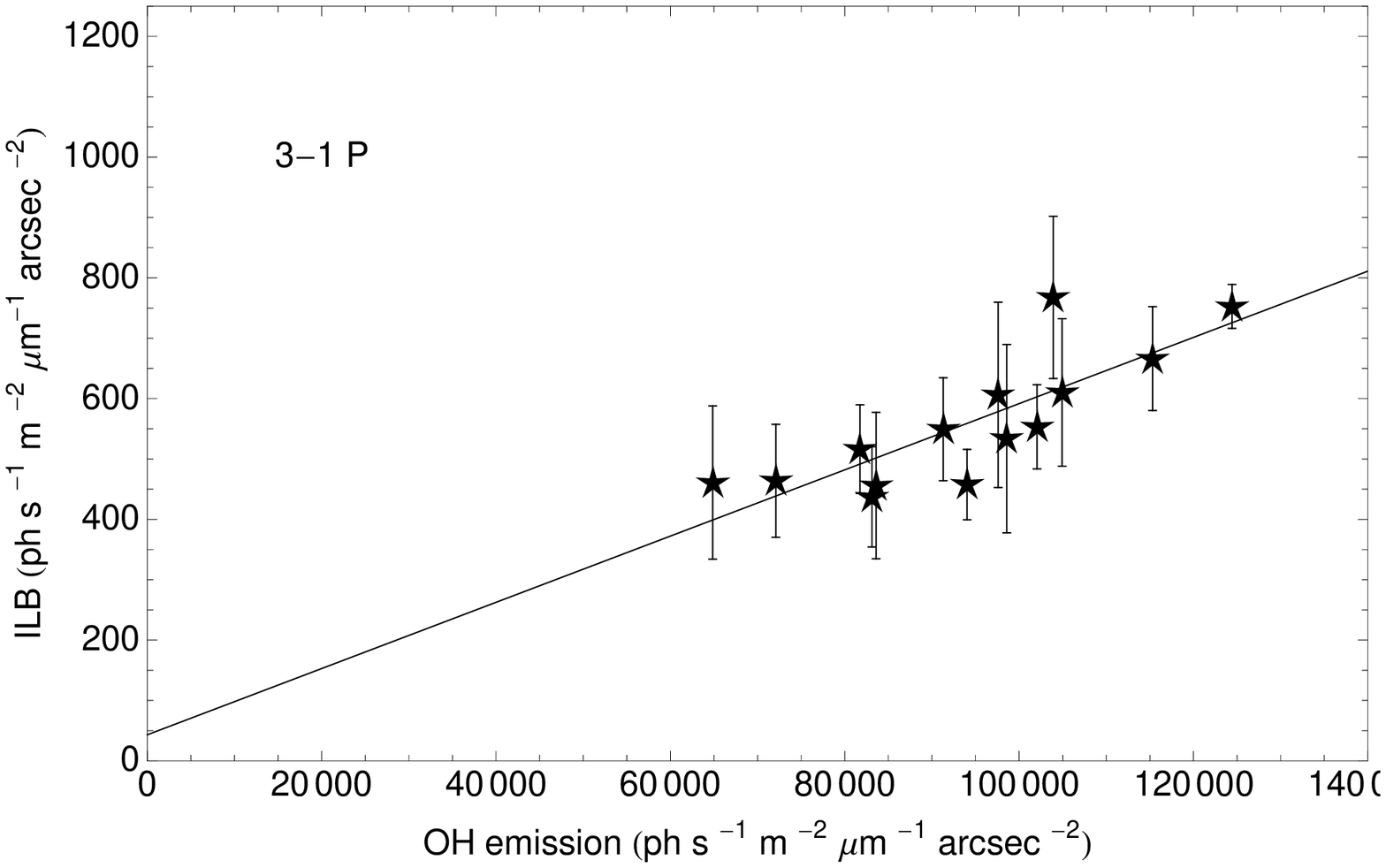} &
\includegraphics[width=0.47\textwidth]{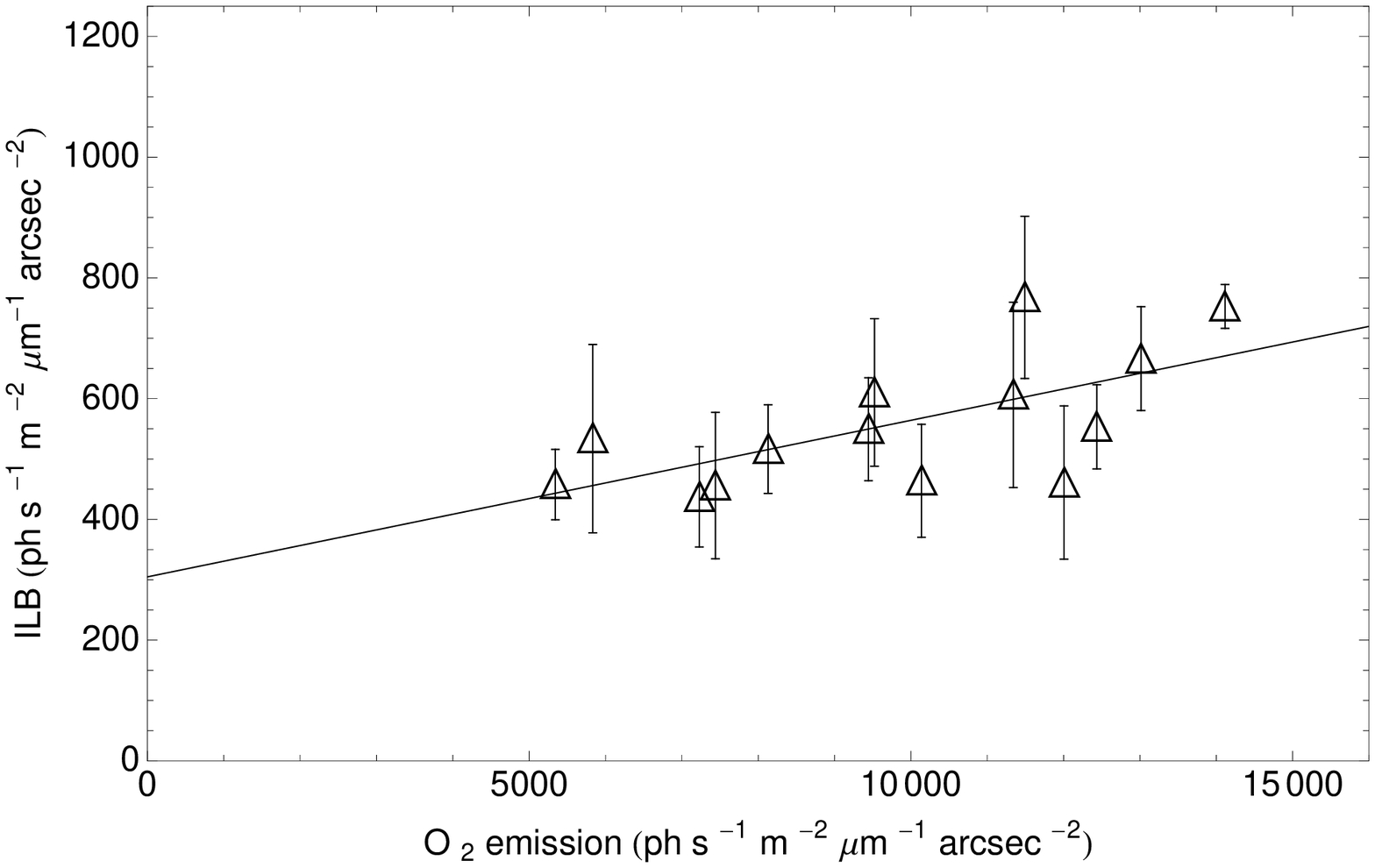} \\
\includegraphics[width=0.47\textwidth]{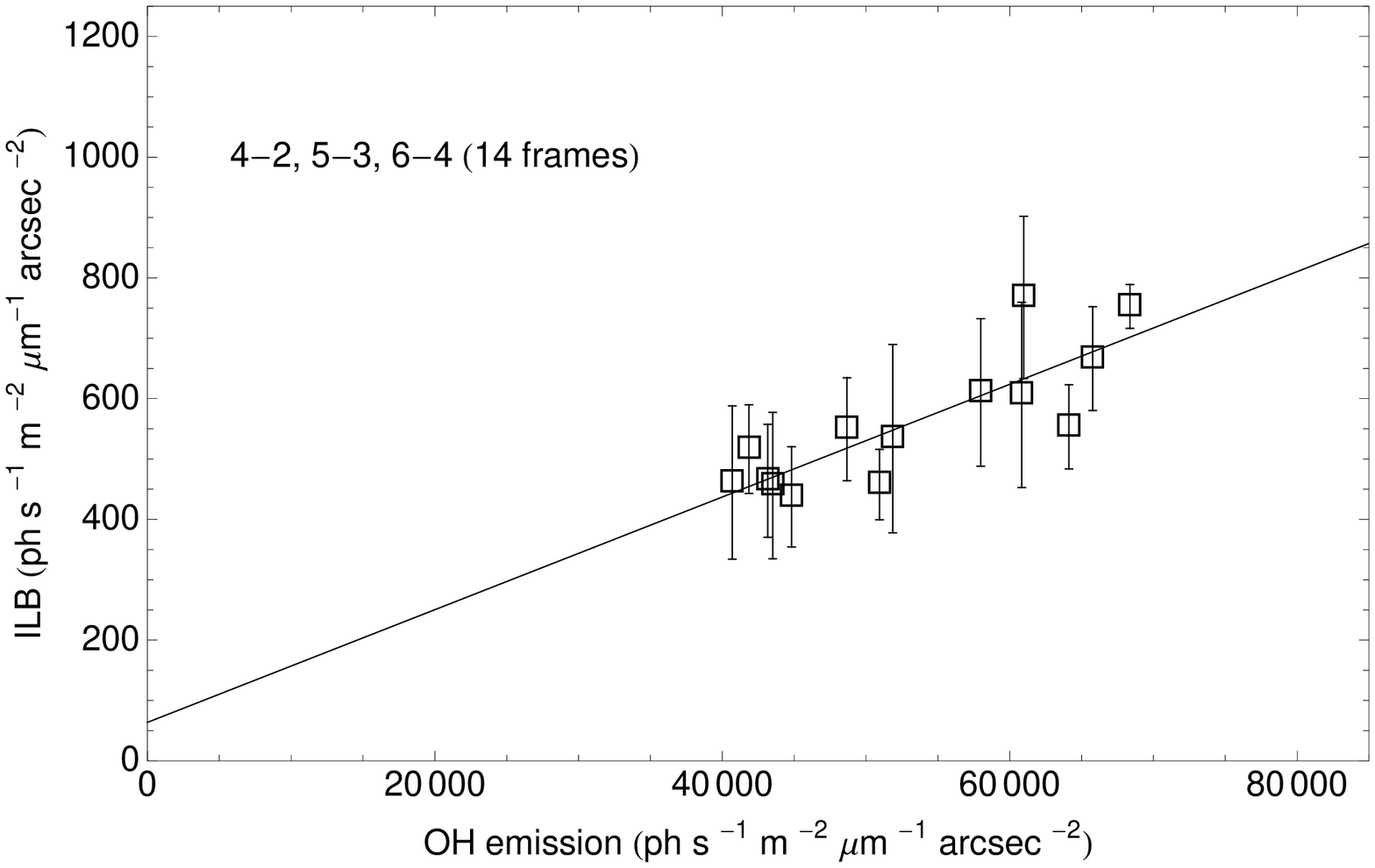} &
\includegraphics[width=0.47\textwidth]{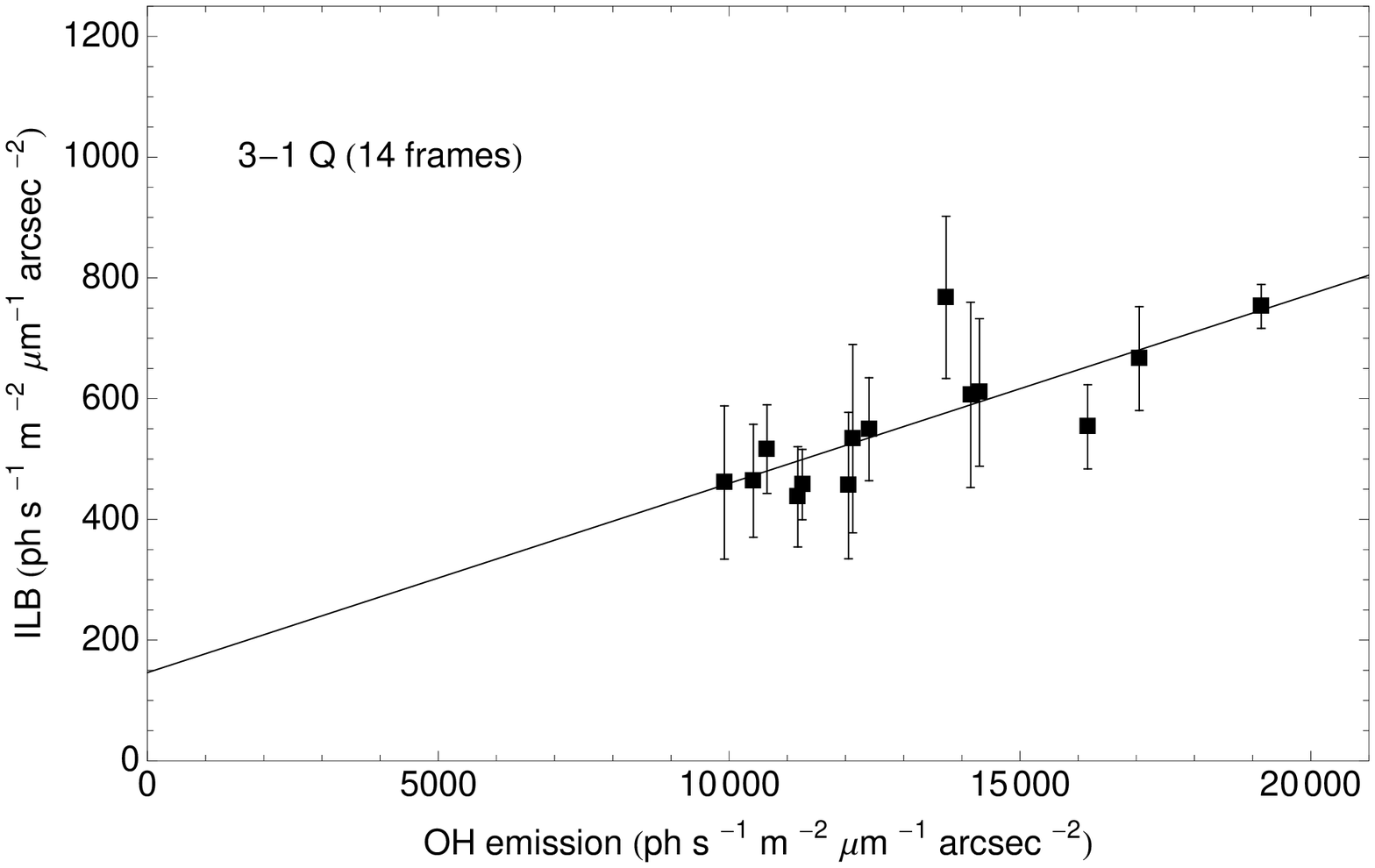} \\
\includegraphics[width=0.47\textwidth]{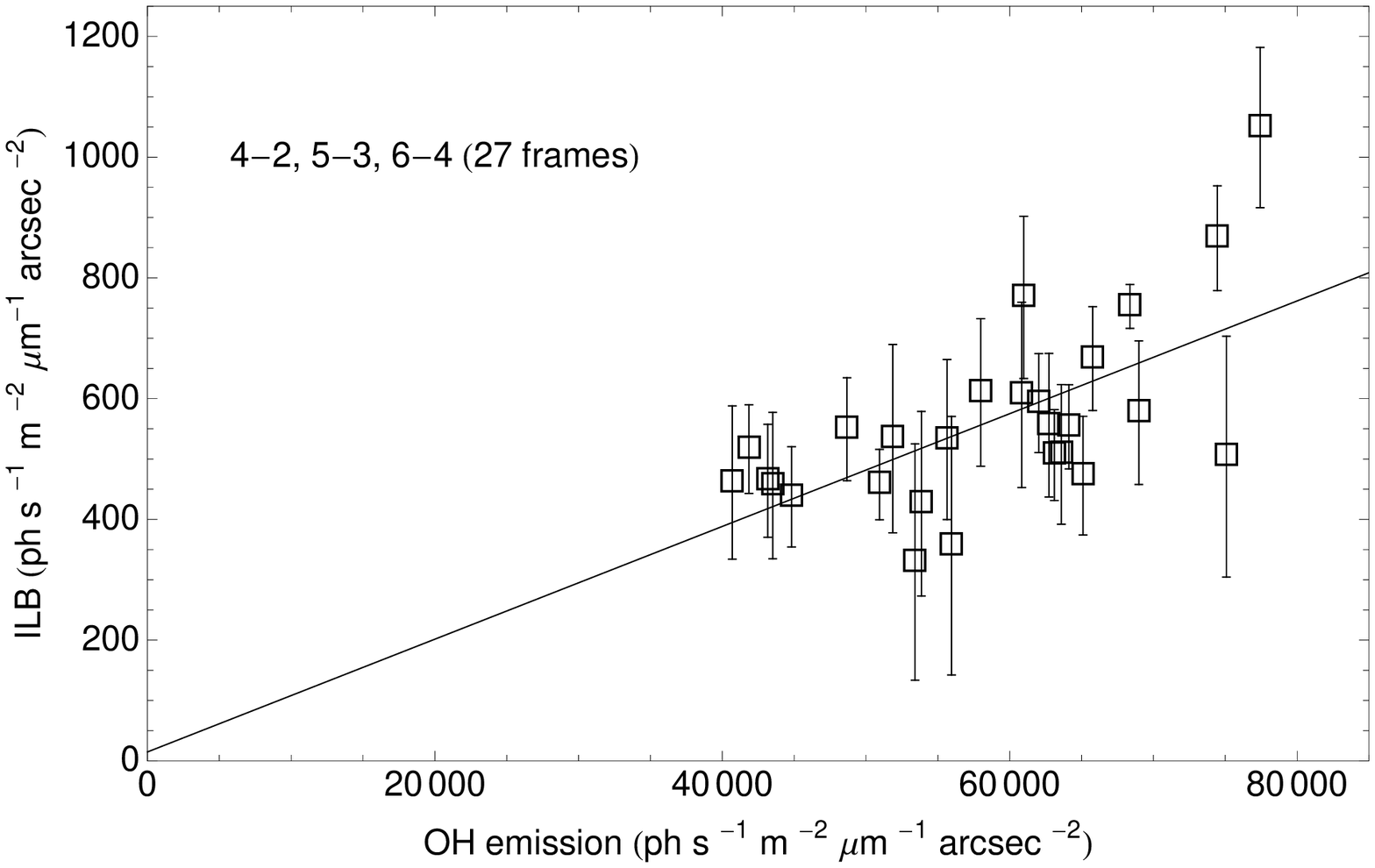} &
\includegraphics[width=0.47\textwidth]{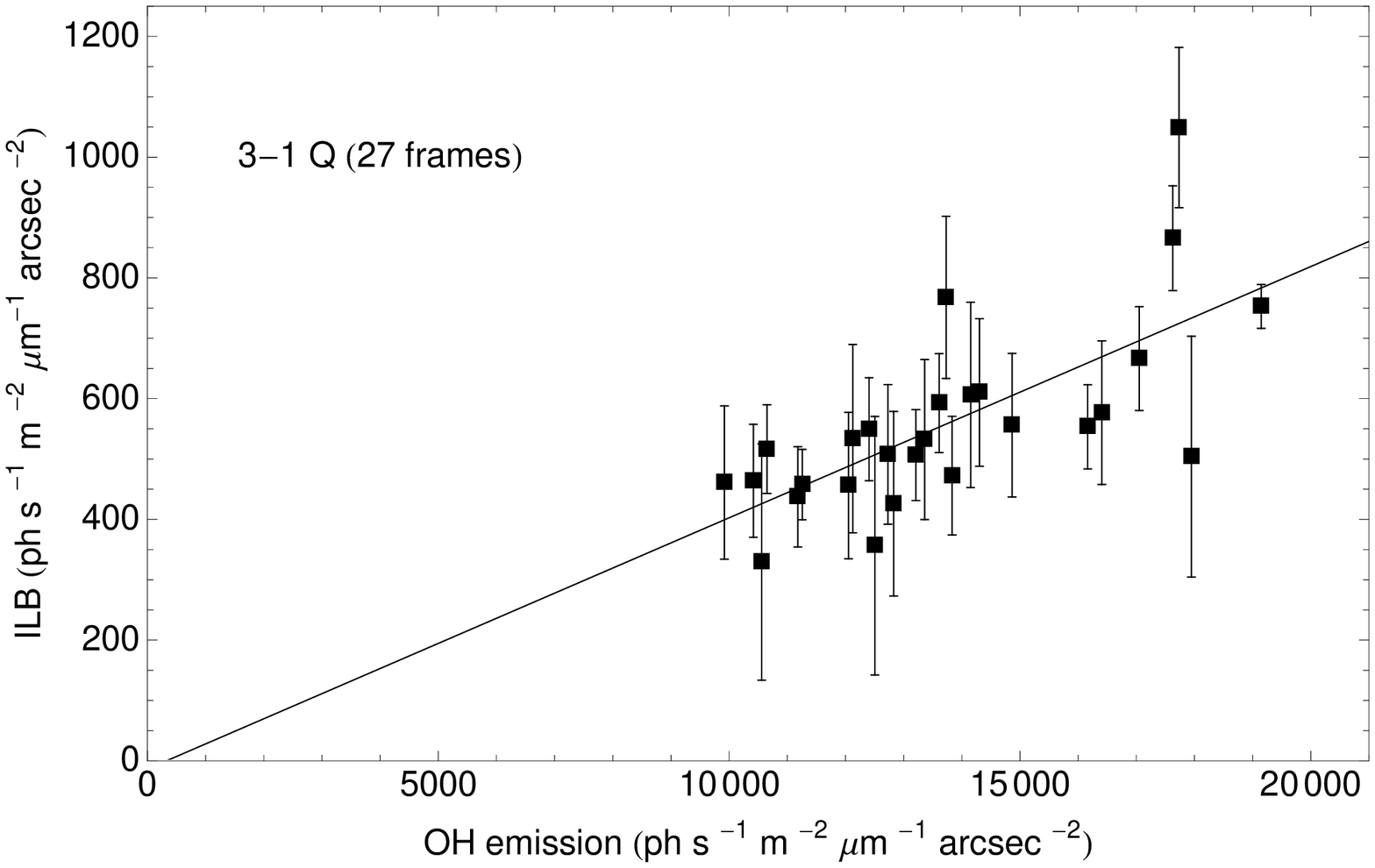} \\
\end{tabular}
\caption{Interline background versus 3--1 Q OH (filled squares), 3--1 P OH (stars), 4--2, 5--3, 6--4 OH (open squares) and O$_{2}$ (triangles) emission for the sky frames of field A. All values have been adjusted to an airmass of unity using  $X_{1}$ for OH and $X_{2}$ for O$_{2}$ and the interline background. The interline background linearly correlates with both OH and O$_{2}$ suggesting that it is OH-like and O$_{2}$-like. However, there are only two sky frames early in the night whether the temporal behaviour of OH and O$_{2}$ are significantly different. Expected contributions from moonlight and ZSL in these frames is very low and should not significantly affect the linear correlation.\label{figure:ILBvsOH}}
\end{figure*}

\subsection{Interline background linear correlations}\label{section:ILBcorrelations}
To investigate further which atmospheric molecules may be responsible for the atmospheric component of the ILB, we examine the linear correlation between the ILB and OH and O$_{2}$ emission. For this purpose, we consider all observations of field A (up to 27 without and 14 with a control fibre) because the night to night variation of OH or O$_{2}$ emission should simply result in a corresponding variation in the ILB if the quantities are correlated. The ZSL contribution to the ILB is nearly identical in all these frames because the ecliptic latitude is nearly the same ($b\approx$\,$-16$\,deg) and this constant value should only affect the best-fitting intercept. The scattered moonlight contribution is small and varies by $\la$\,3\,photons\,s$^{-1}$\,m$^{-2}$\,$\micron^{-1}$\,arcsec$^{-2}$ from frame to frame based on Equation (\ref{equation:ksmodel}) for both Rayleigh and Mie scattering. 

Figure \ref{figure:ILBvsOH} shows the airmass-corrected ILB versus airmass-corrected 3--1 Q OH, 3--1 P OH, 4--2, 5--3, 6--4 OH and O$_{2}$ emission for the sky frames of field A. OH is corrected using $X_{1}$ while O$_{2}$ and the ILB are corrected using $X_{2}$ as before. Again, our choice of $X_{2}$ does not affect any of the results or conclusions. We determine the best-fitting slope and intercept, their standard errors and the coefficient of determination using {\sc LinearModelFit} in {\sc Mathematica}. The best-fitting linear models are shown by the lines in Figure \ref{figure:ILBvsOH} and the fit parameters are listed in Table \ref{table:ILBOH}. 

Considering the 14 of 27 sky frames of field A with a control fibre in order to make a fair comparison of all three sets of OH lines and O$_{2}$, we see a strong linear correlation between OH ($R^{2}\approx$0.650) and a moderate correlation with O$_{2}$ ($R^{2}=0.425$). This result seems to indicate that the ILB is much more OH-like than O$_{2}$-like. However, the temporal behaviour of OH and O$_{2}$ are very similar except very early in the night ($t\la3$\,hr). There are only two sky frames in the entire field A set with $t\la3$\,hr and the correlation analysis does not really distinguish OH and O$_{2}$ in the ILB. Similarly, it is difficult to distinguish between the subtle difference in temporal behaviour between the OH vibrational transitions using correlation analysis. More data helps to distinguish these subtle differences. For 3--1 Q and 4--2, 5--3, 6--4 OH emission, we can double the amount of data by considering all 27 sky frames of field A. Doing so weakens the linear correlation, $R^{2}=0.484$ for 3--1 Q and $R^{2}=0.397$ for 4--2, 5--3, 6--4 OH emission, but these values are similar to the correlation strengths ($R^{2}=0.462$) reported by \citet{ss2012} with a significantly larger data set. When considering all 27 sky frames of field A the ILB correlates with 3--1 Q OH better than 4--2, 5--3, 6--4 OH, as we expected from earlier arguments about the ILB measurement region being near 2--0, 3--1 and 9--6 OH vibrational transitions. 

We conclude that the ILB seems to be instrumentally-scattered OH emission in nature based on its airmass dependence, temporal behaviour and linear correlation with OH. This OH component may be residual from the suppressed 3--1 P lines as suggested by \citet{sce2008}, who show that the even with OH suppression, the ILB should still be dominated by residual OH emission. Alternatively, it may be instrumentally-scattered light from nearby unsuppressed OH lines, such as the 3--1 Q lines. Knowing where the instrumentally-scattered OH comes from is obviously crucial information for perfecting OH suppression fibre technology, but it is difficult to say with the present analysis which OH vibrational transitions are responsible given how similar the vibrational transitions are to each other and the fact that the ILB in the OH suppressed and control spectrum are the same \citep{sce2012}. Additionally, there is possibly a contribution from other atmospheric molecules that dim rapidly early in the night. Beyond this, it is not possible for us to make a definite positive identification of which atmospheric molecules constitute the strong atmospheric component of the ILB. Doing so will require high resolution, high sensitivity spectra of the region where the ILB is measured.

\begin{table}
 \centering
 \begin{minipage}{80mm}
  \caption{Linear correlation for field A\label{table:ILBOH}}
  \begin{tabular}{@{}lccc@{}}
  \hline
   Emission & Slope & Intercept & $R^{2}$ \\
 \hline
 \multicolumn{4}{c}{14 sky frames with control fibre}\\
 \\
3--1 P OH & 0.005\,$\pm$\,0.001 & 0\,$\pm$\,100 & 0.664 \\
3--1 Q OH  & 0.03\,$\pm$\,0.01 & 150\,$\pm$\,100 & 0.620 \\
4--2, 5--3, 6--4 OH & 0.009\,$\pm$\,0.002 & 60\,$\pm$\,100 & 0.692 \\
O$_{2}$ & 0.03\,$\pm$\,0.01 & 300\,$\pm$\,100 & 0.422 \\
\hline
 \multicolumn{4}{c}{27 sky frames without control fibre}\\
 \\
3--1 Q OH  & 0.04\,$\pm$\,0.01 & 0\,$\pm$\,100 & 0.484 \\
4--2, 5--3, 6--4 OH & 0.009\,$\pm$\,0.002 & 0\,$\pm$\,100 & 0.397 \\
\hline
\end{tabular}
\end{minipage}
\end{table}

\subsection{Non-atmospheric contributions}\label{section:ILBnonatmosphere}

Above, we demonstrated that the ILB contains a strong OH-like atmospheric component. Now we turn our attention to non-atmospheric contributions to the ILB, specifically ZSL and scattered moonlight. Based on the estimates from the models shown in Figures \ref{figure:zslmodel} and \ref{figure:moonlightmodel}, these contributions are expected to be very weak compared to the OH-like atmospheric component (mean un-corrected measured ILB $\approx$\,1000\,photons\,s$^{-1}$\,m$^{-2}$\,$\micron^{-1}$\,arcsec$^{-2}$). Nevertheless, following works such as \citet{ss2012}, we search for trends in the ILB with ecliptic latitude, lunar distance, and lunar zenith distance in an attempt to detect the presence of ZSL and scattered moonlight in the ILB. We also attempt to quantify these components in the ILB by a model fitting procedure. Fitting for very weak non-atmospheric components is challenging, especially given the large variability of the atmospheric component. However, the goal of OH suppression is to reduce the atmospheric background component and in future OH suppressed data, we may be able to accurately assess weaker background components by model fitting. 

\subsubsection{Zodiacal scattered light}
First, we attempt to detect the contribution from ZSL in our measured ILB. Of the 45 sky frames in our data set, we have three sky frames taken very close to the Moon ($\rho\la$\,11\,deg), which we refer to as the small $\rho$ set. Judging by eye, there appears to be a significant contribution from scattered moonlight due to Mie scattering by atmospheric aerosols in the small $\rho$ set and we omit it in the analysis of the ZSL component. The remaining 42 sky frames all have $\rho\ga$\,68\,deg and $|\alpha|\ga$\,80\,deg and we refer to these as the large $\rho$ set. According to the extrapolated \citet{ks1991} model, the scattered moonlight component in the large $\rho$ set is $\la$\,2\,phtons\,s$^{-1}$\,m$^{-2}$\,$\micron^{-1}$\,arcsec$^{-2}$ (see Figure \ref{figure:moonlightmodel}). Additionally, \citet{ss2012} find very weak correlations between the ILB and $z_{\mathrm{moon}}$ and $\rho$ in their $H$ band data indicating a very weak scattered moonlight component. Thus, we assume that the ILB in the large $\rho$ set is essentially the strong atmospheric component and ZSL.

\begin{figure}
\includegraphics[width=0.45\textwidth]{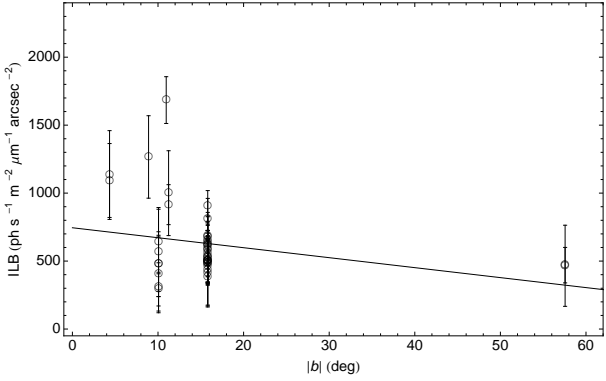}
\caption{Interline background versus ecliptic latitude for 42 sky frames with $\rho\ga$\,68\,deg and $|\alpha|\ga$\,80\,deg. Scattered moonlight component is very small in these frames and the ILB is mostly the atmospheric component (likely OH) and ZSL. Values are corrected for airmass and time dependence assuming an OH component. A very weak correlation ($R^{2}=0.072$) with $b$ is seen indicating a very weak ZSL component compared to the strong atmospheric component.\label{figure:zsl42}}
\end{figure}

Figure \ref{figure:zsl42} shows the ILB versus $|b|$ for the large $\rho$ set. As the ZSL component is quite weak compared to the atmospheric component, we correct for the airmass and time dependence of the atmospheric component by dividing the ILB by $X_{1}$ and $I_{0}(t)$. For $I_{0}(t)$, we use the best-fitting piecewise linear model for 3--1 Q OH emission, listed in Table \ref{table:OHcombined}, as this set of OH lines has the strongest correlation with the ILB for all 27 sky frames of field A. The range of $b$ values is not well-sampled and no obvious trend is visible. The line shows the best-fitting linear model, but the correlation is very weak ($R^{2}=0.072$). We conclude that ZSL is indeed very weak compared to the atmospheric component of the ILB in the $H$ band.

We corroborate this with a model-fitting procedure. Consider the following ILB model including an OH and ZSL component,
\begin{equation}
I_{\mathrm{ILB}}(t,z,b) = I_{0,\mathrm{OH}} \frac{I_{0}(t)X_{1}(z)}{I_{\mathrm{ZSL}}(0)} + I_{0,\mathrm{ZSL}} I_{\mathrm{ZSL}}(b),
\end{equation}
where $I_{0,\mathrm{OH}}$ and $I_{0,\mathrm{ZSL}}$ are free parameters. The OH component has been normalised to the peak value \citet{sce2008} ZSL model such that both $I_{0,\mathrm{OH}}$ and $I_{0,\mathrm{ZSL}}$ are unit-less scaling factors that may be compared directly. We use {\sc NonlinearModelFit} in {\sc Mathematica} to determine the best-fitting parameter values and their standard errors (computed from the covariance matrix) with the constraint that the parameter values be real and positive. The best-fitting ILB model to the large $\rho$ set has $I_{\mathrm{0,OH}}=35,000\,\pm\,10,000$ and $I_{0,\mathrm{ZSL}}=4\,\pm\,3$. The large magnitude and standard error of the OH component are a reflection of the dominance and variability of the OH component, which renders the measurement of the ZSL unreliable in this case. Thus, we adopt the ZSL model of \citet{sce2008} and subtract the expected ZSL contribution from the ILB in our analysis of scattered moonlight.

\begin{figure}
\includegraphics[trim=0 0 0 0,clip,width=0.47\textwidth]{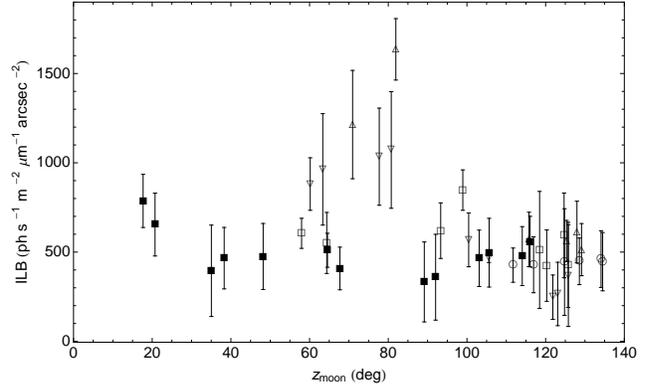}
\caption{Interline background versus lunar zenith distance for 42 sky frames with $\rho\ga$\,68\,deg and $|\alpha|\ga$\,80\,deg. The data is separated by date: 1 Sep (circles), 2 Sep (triangles), 3 Sep (squares), 4 Sep (inverted triangles) and 5 Sep (filled squares). ILB values are ZSL-subtracted and the airmass and time dependence of the large atmospheric component is removed. ILB after moonset are consistently dark. At $z_{\mathrm{moon}}<90$\,deg, the ILB appears to increase with $z_{\mathrm{moon}}$, which is opposite behaviour expected from Equation \ref{equation:ksmodel} and observations of \citet{ss2012}. This trend is unlikely due to scattered moonlight and more likely due to the night to night variation of the strong atmospheric component of the ILB.\label{figure:ILBzmoon42}}
\end{figure}

\subsubsection{Scattered moonlight}\label{section:moonlight}
\citet{ss2012} find very weak dependences on lunar distance ($R^{2}=0.093$) and lunar zenith distance ($R^{2}=0.010$) for the $H$ band ILB and we search for similar correlations in our data. First, we continue our study of the 42 sky frames of the large $\rho$ set. Figure \ref{figure:ILBzmoon42} shows the ILB versus $z_{\mathrm{moon}}$ for these sky frames. The values are ZSL-subtracted using Equation (\ref{equation:zslmodel}) and corrected for airmass and time dependence as before. There is no dependence on $z_{\mathrm{moon}}$ after moonset. There are six data points in the range $60<z_{\mathrm{moon}}<80$\,deg that are a factor of $\approx$\,2 larger than the rest of the data points. The ILB even seems to rise towards $z_{\mathrm{moon}}=90$\,deg, which is opposite to the behaviour expected from Equation (\ref{equation:ksmodel}) and the trend seen in \citet{ss2012}. These sky frames with the ILB value $\ga$\,1000\,photons\,s$^{-1}$\,m$^{-2}$\,$\micron^{-1}$\,arcsec$^{-2}$ are from 2 and 4 September 2011. Within each night, there is no dependence on $z_{\mathrm{moon}}$. Thus, the apparent trend is likely not due to scattered moonlight, but a combination of the small number of data points and the night to night variation of the strong atmospheric component. Thus, our results are in agreement with \citet{ss2012}.

\begin{figure}
\includegraphics[trim=0 0 0 0,clip,width=0.47\textwidth]{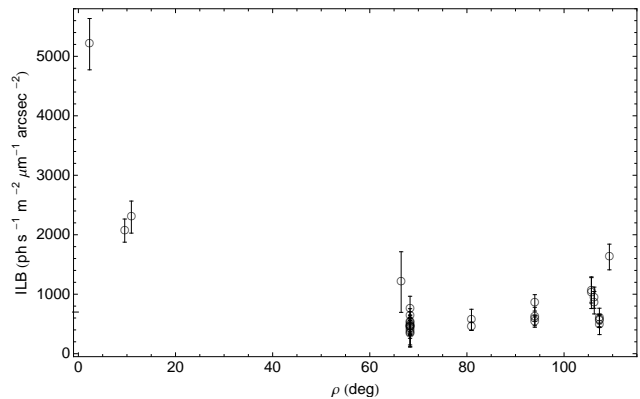}
\caption{Interline background versus lunar distance. The solid and dashed curves correspond to the $I_{\mathrm{moon}}$ curves with $\lambda=1.520\,\micron$, $z_{\mathrm{sky}}=0$\,deg, $z_{\mathrm{moon}}=\rho$, and lunar phase angle $\alpha=-80,-118$\,deg, respectively, scaled by the best-fitting $I_{0,\mathrm{moon}}$ value when fitting the ZSL-subtracted ILB for instrumentally-scattered OH and scattered moonlight. The 3 points at $\la$ 11\,deg contain a significant amount of moonlight ($\ga$ 300\,photon\,s$^{-1}$\,m$^{-2}$\,$\micron^{-1}$\,arcsec$^{-2}$) giving a reliable estimate even amongst the large scatter in the atmospheric component of the ILB.\label{figure:ILBrhoall}}
\end{figure}

Figure \ref{figure:ILBrhoall} shows the ILB versus $\rho$ for the large $\rho$ set and the small $\rho$ set. The ILB values of the small $\rho$ set are ZSL-subtracted but not corrected for the airmass and time dependence of the atmospheric component because the scattered moonlight component in these frames may be quite large. There is no ILB dependence on $\rho$ at large lunar distances. The curve is poorly-sampled between 15 and 60\,deg, but the data show some dependence on lunar distance at $\rho\la$\,11\,deg, which is consistent with Mie scattering by atmospheric aerosols. Note that \citet{ss2012} find a weak correlation with $\rho$, but they do not have any data points within $\approx$\,15\,deg from the Moon. 

Equation (\ref{equation:ksmodel}) predicts $\approx$\,140, 10 and 30\,photons\,s$^{-1}$\,m$^{-2}$\,$\micron^{-1}$\,arcsec$^{-2}$ from moonlight Mie scattered by atmospheric aerosols for sky frames 7, 8 and 32, respectively, in the small $\rho$ set. These contributions are much too small to account for the observed ILB values assuming the only other contributions are from an OH component and ZSL. To see this, consider the following ILB model for the OH and ZSL components,
\begin{equation}
I_{\mathrm{ILB}}(t,z,b) = I_{0,\mathrm{OH}} I_{0}(t)X_{1}(z) + I_{\mathrm{ZSL}}(b),
\end{equation}
where the first term is the time-dependent, airmass-dependent OH component with $I_{0,\mathrm{OH}}$ as a free parameter and the second term is the ZSL model of \citet{sce2008}.  As before, we use {\sc NonlinearModelFit} in {\sc Mathematica} to determine the best-fitting $I_{0,\mathrm{OH}}$ value to the un-corrected ILB of the large $\rho$ set with the constraint that the value be real and positive. Then, we use this best-fitting ILB model to estimate the OH+ZSL contribution in the small $\rho$ set. To this estimate, we include a conservative uncertainty estimate of $\sigma=50$ per cent to account for the large variability of OH. Then, we estimate the scattered moonlight contribution by ILB$_{\mathrm{measured}}-I_{\mathrm{ILB}}(t,z,b)(1+\sigma)$. This yields a scattered moonlight contribution of at least 3500, 580 and 880\,photons\,s$^{-1}$\,m$^{-2}$\,$\micron^{-1}$\,arcsec$^{-2}$ for sky frames 7, 8 and 32, respectively. Compared to the estimate from Equation (\ref{equation:ksmodel}), the contribution from Mie scattered moonlight is $\approx$\,25, 60 and 28 times greater than expected.    

This result is not too surprising, considering our estimates using Equation (\ref{equation:ksmodel}) utilise the empirically measured Mie scattering function over Mauna Kea. Differences in the aerosol composition and concentration, which is affected by factors such as industrial pollution, volcanic eruptions and sand storms, between Siding Spring and Mauna Kea could certainly account for the discrepancy between the model and observations. Given its location and low altitude, Siding Spring is likely more susceptible to aerosol scattering making scattered moonlight a potentially more significant contributor to the NIR background at this location. However, more night sky observations at smaller lunar phase angles ($|\alpha|\la$\,30\,deg) and at low lunar distances ($\rho\la$\,30\,deg) where the scattered moonlight component is comparable to the atmospheric component are necessary to firmly establish the impact aerosol scattering of moonlight on NIR observations at Siding Spring.

\subsection{Absolute interline background}
Previously, we used the variation of the ILB measured at 1.520\,$\micron$ with airmass, time after sunset, ecliptic latitude, lunar zenith distance and lunar distance to determine the presence of non-thermal atmospheric emission, ZSL and scattered moonlight in the ILB when using OH suppression fibres. We established that the ILB is dominated an OH-like atmospheric source. The end goal of OH suppression is to reduce the ILB for higher sensitivity observations. While the current generation of OH suppression have yet to realise this ILB reduction \citep{sce2012,cqt2013}, we measure the absolute sky brightness of the ILB with the current observations under the darkest conditions to establish a benchmark for future OH suppressed observations. We select nine low airmass ($X_{2}<1.2$), middle of the night ($5<t<9$\,hr after sunset) and moonless ($z_{\mathrm{moon}}>90$\,deg) sky frames for this dark ILB measurement, which yields an ILB of 560\,$\pm$\,120\,photons\,s$^{-1}$\,m$^{-2}$\,$\micron^{-1}$\,arcsec$^{-2}$ from the mean and standard deviation. 

Our measured dark ILB is similar to the values of \citet{ss2012} and \citet{maihara1993} at 1.665\,$\micron$ at $R=6000$ and $R=17,000$, respectively, without OH suppression. This suggests that the ILB in their spectra and ours is the same, hence the suppression of the 103 brightest OH lines is not affecting the ILB as modelled by \citet{sce2008} and the ILB is dominated by emission from non-suppressed atmospheric emission (OH or other). If this is the case, it would explain the nearly identical ILB in the OH suppressed and non-suppressed sky spectra taken with GNOSIS+IRIS2. However, we remind the reader that these observations are detector noise-dominated and it is possible that there is an ILB reduction in our OH suppressed spectrum that we are unable to observe among the detector noise. In any case, it is difficult the form a compelling argument at this point considering we measure the ILB at a different wavelength. \citet{sce2012} measure a value of 860\,$\pm$\,210\,photons\,s$^{-1}$\,m$^{-2}$\,$\micron^{-1}$\,arcsec$^{-2}$ over the entire $H$ band with OH suppression using GNOSIS+IRIS2 data indicating that our ILB measurement window may be one of the darker windows in the $H$ band. With the current GNOSIS+IRIS2 data, we are unable to reliably measure the ILB at 1.665\,$\micron$ due to contamination by instrumental thermal emission and the large uncertainties associated with the removal of the thermal background. A direct comparison at 1.665\,$\micron$ with OH suppression fibres in a regime that is sky background-limited using an optimised and cooled spectrograph is left for future work \citep{horton2012}. 

\section{Summary}\label{section:summary}
In this paper we analysed 19\,hr of $H$ band observations taken with GNOSIS, the first OH suppression unit to utilise OH suppression fibres with fibre Bragg gratings and the IRIS2 spectrograph at the AAT. These data include 45 night sky observations covering a range of sky positions and lunar conditions. From these data, we examined the spatial and temporal behaviours of the interline background, OH and O$_{2}$ emission to determine the nature of the ILB measured at 1.520\,$\micron$ when using the first generation of OH suppression fibres with a $R\approx2400$ spectrograph. This analysis is a critical step in the development of OH suppression fibre technology, which may potentially impact observational astronomy at the level of adaptive optics or greater. 

The region where the ILB was measured is populated exclusively by OH lines corresponding to 2--0, 3--1 and 9--6 vibrational transitions and these are the most likely to directly contribute instrumentally-scattered OH into the measured ILB. Thus, we studied the temporal behaviour of OH lines corresponding to 3--1 and 4--2, 5--3 and 6--4 vibrational transitions measured in the OH suppressed and non-suppressed sky spectra when available. We found that the temporal behaviour of OH emission is best modelled by a gradual linear decrease the first half of the night followed by a gradual linear rise during the second half of the night following the behaviour of the solar elevation angle. O$_{2}$ emission on the other hand, exhibits very rapid dimming after sunset. 

For space-based observations, the dominant NIR background component is zodiacal scattered light. Unfortunately, given the large variability of the atmospheric component in the ILB and the relative weakness of the ZSL component, we were unable to reliably measure ZSL in our data. We adopted the ZSL model of \citet{sce2008} and removed this component from the ILB where appropriate.


The ILB contribution from scattered moonlight in the $H$ band is undetectable, except at small lunar distances ($\rho\la$\,11\,deg) where Mie scattering by atmospheric aerosols is very efficient. This contribution is much larger than expected from a model based on $V$ band measurements at Mauna Kea by \citet{ks1991} and extrapolated to the $H$ band using the typical wavelength dependences for the scattering efficiency of Rayleigh and Mie scattering. Due to its location and altitude, Siding Spring is likely more susceptible to aerosol scattering by industrial pollution and sand storms compared to Mauna Kea. Thus, the scattered moonlight contribution to the NIR ILB at small lunar distance is potentially more significant at Siding Spring compared to other sites. We have found evidence for this, but more measurements of the night sky background near full Moon and at small lunar distance are necessary to firmly establish the significance of aerosol scattering of moonlight at Siding Spring. Otherwise, for large lunar distances ($\rho\ga$\,60\,deg) and large lunar phase angles ($|\alpha|\ga$\,80\,deg), we found no dependence on lunar distance or lunar elevation angle indicating a weak contribution from moonlight in the $H$ band in agreement with the results of \citet{ss2012} using data from FIRE at Magellan. 



Although moonlight dominates the ILB at very small lunar distances the background when using OH suppression fibres is otherwise dominated by non-thermal atmospheric emission, potentially either line emission scattered by the instrument or an airglow continuum. The ILB exhibits the characteristic $X[z(t)]$ behaviour expected of emission from an atmospheric molecule when following the same field throughout the night. Additionally, the temporal behaviour of the ILB is similar to that of OH and O$_{2}$ with the dimming of the ILB early in the night being intermediate between the slow rate of OH and the fast rate of O$_{2}$. This suggests that the ILB contains atmospheric emission from multiple molecules with both slow dimming rates like OH and rapid dimming rates like O$_{2}$. There are no O$_{2}$ transitions in the ILB measurement region but O$_{2}$-like molecules may be responsible for the intermediate dimming rate of the ILB early in the night. There is also a strong linear correlation between the ILB and OH emission, which strengthens the case that the atmospheric component of the ILB is dominated by instrumentally-scattered OH. However, it is not possible to distinguish which lines or vibrational transitions are responsible given their similar long-term temporal behaviours. 

The absolute ILB at 1.520\,$\micron$ under dark conditions with OH suppression is 560\,$\pm$\,120\,photons\,s$^{-1}$\,m$^{-2}$\,$\micron^{-1}$\,arcsec$^{-2}$. This value is similar to previous non-suppressed measurements and our own measurement from a control fibre suggesting that the suppression with the current grating design does not affect the ILB and it is dominated by emission from non-suppressed atmospheric emission. There is some uncertainty regarding this conclusion given that the current GNOSIS+IRIS2 observations are detector noise-dominated and the ILB comparison was made at two different wavelengths. We are currently designing an optimised and cooled fibre-fed spectrograph, called PRAXIS, with a high performance, low noise 1.7\,$\micron$ cutoff Hawaii-2RG detector to be used with the GNOSIS for further development of OH suppression fibre technology. GNOSIS+PRAXIS will provide sky background-limited data with significantly reduced thermal noise \citep{horton2012}. If no ILB reduction is seen in these data, then it will be clear that non-suppressed atmospheric emission is responsible for the ILB. In that case, an ILB reduction may still be possible with improved fibre design, i.e. suppressing more sky lines. 

For example, there is evidence that non-suppressed OH lines originally thought to be very weak are responsible for the lack of ILB reduction. This could certainly account for the observed linear correlation between the ILB and OH. The simulations of \citet{sce2008} show an ILB reduction when the sky is modelled according to the OH line list of \citet{rousselot2000}. Here, only the 103 brightest OH doublets need to be suppressed because the other non-suppressed lines are so weak that they do not matter. However, the OH line list of \citet{abrams1994} indicates that some of the non-suppressed lines are much stronger scattering more light into the interline regions. Comparing the observed GNOSIS+IRIS2 spectrum to a simulated OH suppressed spectrum using the OH line spectrum of \citet{rousselot2000} and a simulated spectrum using an updated OH line spectrum consisting of an amalgamation of \citet{abrams1994} and \citet{rousselot2000} and giving precedence to \citet{abrams1994} when a line occurs in both lists, indicates that the latter is a better fit to the observations. If these non-suppressed lines are stronger and responsible for the lack of ILB reduction, more accurate model of the night sky should be used to determine the number of lines that need to be suppressed to achieve the desired ILB reduction. This highlights the fact that for proper sky suppression, accurate sky line positions and strengths are critical. New measurements of the OH lines are being carried out with high resolution spectra taken with CRIRES (CRyogenic high-resolution InfraRed Echelle Spectrograph) at the Very Large Telescope (Chris Lidman, private communication). Future grating designs will incorporate these new measurements in order to optimise the performance of OH suppression fibres for deep ground-based NIR spectroscopy.

\section*{Acknowledgments}
The GNOSIS team acknowledges funding by ARC LIEF grant LE100100164. CQT gratefully acknowledges support by the National Science Foundation Graduate Research Fellowship under Grant No. DGE-1035963. The authors thank the referee for a very thorough report that has greatly improved this manuscript. CQT thanks Robert Content, James Graham, Saul Perlmutter and Andrew Sheinis for helpful comments and suggestions. CQT thanks Christopher Betters, Julia Bryant, Seong-sik Min, Emma Simpson, Ren\'ee Pelton and Billy Robbins for assistance with some of the technical aspects of this work.

\label{lastpage}

\end{document}